\documentclass[amsmath,amssymb,showkeys,superscriptaddress,aps,prd,twocolumn]{revtex4}
\usepackage{graphicx}
\usepackage{pdfpages}
\usepackage[caption=false]{subfig}
\usepackage{multirow}
\usepackage{appendix}
\usepackage{graphicx,epstopdf}
\usepackage{subfig}

\def\xslash{x\!\!\!\slash }

\usepackage{colortbl}
\usepackage{xcolor}
\usepackage[colorlinks=true, urlcolor=blue, linkcolor=blue, citecolor=blue]{hyperref}

\begin{document}

\title{Magnetic dipole moments of the $Z_{c}(4020)^+$, $Z_{c}(4200)^+$, $Z_{cs}(4000)^{+}$ and  $Z_{cs}(4220)^{+}$ states in light-cone QCD}

\author{Ula\c{s}~\"{O}zdem}%
\email[]{ulasozdem@aydin.edu.tr (corresponding author)}
\affiliation{ Health Services Vocational School of Higher Education, Istanbul Aydin University, Sefakoy-Kucukcekmece, 34295 Istanbul, Turkey}
\author{Ay\c{s}e Karadeniz~Y{\i}ld{\i}r{\i}m }%
\email[]{aysekaradeniz@aydin.edu.tr}
\affiliation{ Health Services Vocational School of Higher Education, Istanbul Aydin University, Sefakoy-Kucukcekmece, 34295 Istanbul, Turkey}

\date{\today}
 
\begin{abstract}
The magnetic dipole moments of the $Z_{c}(4020)^+$, $Z_{c}(4200)^+$, $Z_{cs}(4000)^{+}$ and  $Z_{cs}(4220)^{+}$ states are extracted in the framework of the light-cone QCD sum rules.
In the calculations, we use the hadronic molecular form of interpolating currents, and photon distribution amplitudes to get the magnetic dipole moment of $Z_{c}(4020)^+$, $Z_{c}(4200)^+$, $Z_{cs}(4000)^{+}$ and  $Z_{cs}(4220)^{+}$ tetraquark states.
The magnetic dipole moments are obtained as 
 $\mu_{Z_{c}} = 0.66^{+0.27}_{-0.25}$, 
 $\mu_{Z^{1}_{c}}=1.03^{+0.32}_{-0.29}$,
 $\mu_{Z_{cs}}=0.73^{+0.28}_{-0.26}$, 
$\mu_{Z^1_{cs}}=0.77^{+0.27}_{-0.25}$ 
 for the  $Z_{c}(4020)^+$, $Z_{c}(4200)^+$, $Z_{cs}(4000)^{+}$ and  $Z_{cs}(4220)^{+}$ states, respectively. 
We observe that the results obtained for the $Z_{c}(4020)^+$, $Z_{c}(4200)^+$, $Z_{cs}(4000)^{+}$ and  $Z_{cs}(4220)^{+}$  states are large enough to be measured experimentally. 
As a by product, we predict the  magnetic dipole moments of the neutral $Z_{cs}(4000)$ and  $Z_{cs}(4220)$ states.
The results presented here can serve to be helpful knowledge in experimental as well as theoretical studies of the properties of hidden-charm tetraquark states with and without strangeness.
\end{abstract}

\maketitle

\section{Introduction}
The standard hadrons, made of the strongly interacting quarks and gluons, are mesons and baryons. There have been proposed particles having configurations out of the standard hadrons and some of them have been observed in experiments. These are exotic particles including XYZ particles, hybrids, meson molecules, tetraquarks, pentaquarks etc. X(3872) is the first four quark structure which contains charm quark, and it was experimentally observed in 2003 by Belle. After X(3872), charged or uncharged, many exotic hadrons were observed. 
Until now, there are six members in the group of the charged charmed exotic resonances: $Z_c(3900)$, $Z_c(4020)$, $Z_c(4050)$,  $Z_c(4200)$, $Z_2(4250)$ and  $Z_c(4430)$ reported in decays into final states contain a pair of light and charm quarks  \cite{Choi:2007wga,Aaij:2014jqa,Mizuk:2008me,Ablikim:2013mio,Liu:2013dau,Ablikim:2013wzq,Ablikim:2013emm,Chilikin:2014bkk,Wang:2014hta,Collaboration:2011gja}.
They cannot be classified as traditional charmonium states because of their electric charge, they must be exotic states with a minimum quark content $ c \bar {c} u \bar {d} $/$ c \bar {c} d \bar {u} $.
Many attempts have been made to figure out the substructure of these exotic states, but the nature of most of them is still unclear.
As tetraquark states with quark contents $ c \bar {c} u \bar {d} $/$ c \bar {c} d \bar {u} $,  family of  these charged exotic states were generally studied as diquark-antidiquark and molecular pictures~\cite{Faccini:2012pj,Esposito:2014rxa,Chen:2016qju,Ali:2017jda,Esposito:2016noz,Olsen:2017bmm,Lebed:2016hpi,Guo:2017jvc,Nielsen:2009uh,Brambilla:2019esw,Liu:2019zoy, Agaev:2020zad, Dong:2021juy,Chen:2015ata}.

Very recently, three charged hidden-charm tetraquark states with strangeness  were reported by the BESIII~\cite{Ablikim:2020hsk} and LHCb~\cite{Aaij:2021ivw} Collaborations. 
The spin-parity of these states are assumed to favor $J^P = 1^+$ and their  quark composition is most likely  $ c \bar c s\bar u $/$ c \bar c \bar s u $.  
Their masses and  widths are
\begin{align}
 Z_{cs}(3985): \mbox{M} &= 3982.5\pm 2.1^{+1.8}_{-2.6}~\mbox{MeV}, ~~~~~~ \nonumber\\
   \Gamma&=12.8^{+5.3}_{-4.4}\pm 3.0~\mbox{MeV},\\
   Z_{cs}(4000): \mbox{M} &= 4003.0\pm 6.0^{+4.0}_{-14.0}~\mbox{MeV}, ~~~~~ \nonumber\\
   \Gamma &=131.0 \pm 15 \pm 26~\mbox{MeV},\\
   Z_{cs}(4220): \mbox{M} &= 4216.0 \pm 24.0^{+43}_{-30}~\mbox{MeV}, ~~~~~\nonumber\\
   \Gamma &=233.0 \pm 52^{+97}_{-73}~\mbox{MeV}.
\end{align}
The $Z_{cs}(3985)$ and $Z_{cs}(4000)$ states have very different decay widths and can be considered as two different states, in spite of being very close in mass. This important difference means that $Z_{cs}(3985)$ and $Z_{cs} (4000)$ can have different origins.
However, there is a possibility that both states are the same, but their masses and  decay widths may not be consistent because of experimental resolution and/or coupled-channels effects.
Moreover, the $ Z_{cs}(3985) $ and $ Z_{cs}(4220) $ particles can be interpreted as strange partners of the  $ Z_{c}(3900) $ and $Z_{c}(4020)$ states, respectively.
Clarification of the aforementioned possibilities regarding these states is among the questions that need to be answered.
These discoveries triggered interesting theoretical examinations in the context of various approaches aimed to account for substructure of the $ Z_{cs}(3985) $,  $Z_{cs}(4000)$ and $Z_{cs}(4220)$ states and calculate their parameters~\cite{Liu:2020nge,
Meng:2020ihj,Wang:2020kej,Chen:2020yvq,Cao:2020cfx,Du:2020vwb,Sun:2020hjw,Wang:2020rcx,Wang:2020htx,Wang:2020iqt,Azizi:2020zyq,Jin:2020yjn,Simonov:2020ozp,Sungu:2020zvk,Ikeno:2020csu,Guo:2020vmu,Zhu:2021vtd,Wang:2020dgr,Ge:2021sdq,Maiani:2021tri,Ortega:2021enc,Yang:2020nrt,Chen:2021erj,Meng:2021rdg}. 
Most of these studies are related to the calculation of spectroscopic parameters.

The electromagnetic properties like masses and decays are the important parameters of a hadron, which are measurable and computable. Therefore, the examination of these properties is important to understand their substructure, non-perturbative picture and dynamics of the exotic states.
In the present study, we calculate the magnetic dipole moments (MDMs) of the $Z_{c}(4020)^+$, $Z_{c}(4200)^+$, $Z_{cs}(4000)^{+}$ and  $Z_{cs}(4220)^{+}$ (hereafter we
will show these states as  $Z_{c}$, $Z^1_{c}$, $Z_{cs}$ and $Z^1_{cs}$, respectively) tetraquark states by considering them as the  molecular picture in the framework of light-cone QCD sum rules (LCSR). 
The LCSR is a powerful method in investigating the exotic hadron characteristics and have been performed successfully to compute the masses, decay constants,  magnetic moments, etc. 
In the LCSR, the operator product expansion (OPE) is carried out
over twist near the light cone. In this case, there come into view matrix elements of non-local operators between vacuum and  photon state. These matrix elements are expressed with respect to the distribution amplitudes (DAs) of the photon~\cite{Chernyak:1990ag, Braun:1988qv, Balitsky:1989ry}.
There are a few studies in the literature where the MDMs of the hidden-charm tetraquark states were extracted~\cite{Ozdem:2021yvo,Ozdem:2017jqh,Xu:2020qtg,Wang:2017dce,Xu:2020evn}. 
%
%
 
 The outline of our paper is as follows. In sec. \ref{secII}, we acquire the QCD sum rules for the MDMs of the tetraquark diquark-antidiquark and molecular states. Sec. \ref{secIII} is devoted to the numerical analysis and discussion on the obtained results. 
 The obtained results are summarized and discussed in Sec. \ref{secIV}. The explicit expressions of the magnetic moment of the $Z_{c(s)}$ tetraquark states are given in Appendix.

 \begin{widetext}
  
\section{Formalism}\label{secII}

It is well known that in the LCSR approach, the correlation function  is obtained in two different ways: Hadronic and QCD sides.  In order to obtain the expression from the hadronic side, the correlation function is saturated with the same quantum numbers as the interpolating currents.   In the QCD side, the correlation function is obtained in deep Euclidean region, $q^2 << 0$, with the help of the OPE.  As usual, in the LCSR approach,  in order  to suppress the contribution of the continuum and  higher states  Borel transformation and continuum subtractions are applied. The sum rules for the MDMs of the $Z_{c(s)}$ states are obtained equating both representations of the correlation function  via the quark-hadron duality ansatz.

We started with two point correlation function to obtain the sum rules for the MDMs of $Z_{c(s)}$ which can be written as
\begin{equation}
 \label{edmn01}
\Pi _{\mu \nu }(p,q)=i\int d^{4}xe^{ip\cdot x}\langle 0|\mathcal{T}\{J_{\mu}(x)
J_{\nu }^{\dagger }(0)\}|0\rangle_{\gamma}, 
\end{equation}%
where  $\gamma$ is the external  electromagnetic field and  $J_{\mu(\nu)}(x)$ is the interpolating current of the $Z_{c(s)}$ states. In the  molecular  picture with quantum numbers $J^{P}=1^{+}$, it is given as 
\begin{align}
\label{curr}
    J_{\mu}^{Z_{c}}(x)&=\frac{1}{\sqrt{2}}\Big\{[ \bar d_a(x) \gamma^{\alpha} c_a(x)][\bar c_b(x) \sigma_{\alpha\mu}\gamma_5 u_b(x)] 
-[\bar d_a(x) \sigma_{\alpha\mu}\gamma_5 c_a(x)][\bar c_b(x) \gamma^{\alpha} u_b(x)]\Big\},\\
 \label{curr2}
 J_{\mu}^{Z^1_c}(x)&=\frac{1}{\sqrt{2}}\Big\{[ \bar d_a(x) \gamma_5 c_a(x)][\bar c_b(x) \gamma_\mu u_b(x)]
+[\bar d_a(x) \gamma_\mu c_a(x)][\bar c_b(x) \gamma_5 u_b(x)]\Big\},\\
%
 \label{curr3}
 J_{\mu}^{Z_{cs}}(x)&=\frac{1}{\sqrt{2}}\Big\{[ \bar s_a(x) \gamma_5 c_a(x)][\bar c_b(x) \gamma_\mu u_b(x)]
+ [\bar s_a(x) \gamma_\mu c_a(x)][\bar c_b(x) \gamma_5 u_b(x)]\Big\},\\
%
\label{curr4}
   J_{\mu}^{Z^1_{cs}}(x)&=\frac{1}{\sqrt{2}}\Big\{[ \bar s_a(x) \gamma^{\alpha} c_a(x)][\bar c_b(x) \sigma_{\alpha\mu} \gamma_5 u_b(x)]
-[\bar s_a(x) \sigma_{\alpha\mu}\gamma_5 c_a(x)][\bar c_b(x) \gamma^{\alpha} u_b(x)]\Big\},
  \end{align}
where $\sigma_{\mu\nu}=\frac{i}{2}[\gamma_{\mu},\gamma_{\nu}]$. 
The MDMs of the neutral $Z_{cs}$ and $Z^1_{cs}$ states will also be calculated. To do this, we substitute the d-quark instead of the u-quark in Eqs. (\ref{curr3}) and (\ref{curr4}). For the sake of brevity, only the numerical results of the  neutral $Z_{cs}$ states will be presented. It is obvious that MDMs are zero for the neutral $ Z_c $ and $ Z^1_c $ states.

The correlation function  in Eq. (\ref{edmn01}) can be calculated in terms of quark and gluon degrees of freedom. To do this, we substitute interpolating currents given in Eqs. (\ref{curr}) to (\ref{curr4})   into the correlation function and  contract the corresponding quark fields via the Wick's theorem. 
As a results, we get following expressions for the $Z_{c(s)}$ states

\begin{align}
\label{neweq}
\Pi _{\mu \nu }^{\mathrm{QCD-Z_{c}}}(p,q)&=\frac{i}{2}
\int d^{4}xe^{ipx} \langle 0 | \Big\{ 
\mathrm{Tr}\Big[\gamma^{\alpha}{S}_{c}^{aa^{\prime }}(x)\gamma ^{\beta}S_{d}^{a^{\prime }a}(-x)\Big] 
\mathrm{Tr}\Big[\sigma_{\mu\alpha}\gamma _{5 }{S}_{u}^{bb^{\prime }}(x)\gamma _{5}\sigma_{\nu\beta}S_{c}^{b^{\prime }b}(-x)\Big] \notag \\
&-\mathrm{Tr}\Big[ \gamma^{\alpha}{S}_{c}^{aa^{\prime}}(x)\gamma _{5}\sigma_{\nu\beta}S_{d}^{a^{\prime }a}(-x)\Big]   
\mathrm{Tr}\Big[ \sigma_{\mu\alpha}\gamma_{5 }{S}_{u}^{bb^{\prime }}(x)\gamma^{\beta}S_{c}^{b^{\prime }b}(-x)\Big] \notag \\
&-\mathrm{Tr}\Big[\sigma_{\mu\alpha}\gamma _{5}{S}_{c}^{aa^{\prime }}(x)\gamma^{\beta }S_{d}^{a^{\prime}a}(-x)\Big]    
\mathrm{Tr}\Big[ \gamma^{\alpha}{S}_{u}^{bb^{\prime}}(x)\gamma_{5}\sigma_{\nu\beta}S_{c}^{b^{\prime }b}(-x)\Big] \notag \\
&+\mathrm{Tr}\Big[\sigma_{\mu\alpha}\gamma_{5 }{S}_{c}^{aa^{\prime }}(x)\gamma _{5}\sigma_{\nu\beta}S_{d}^{a^{\prime}a}(-x)\Big]  
\mathrm{Tr}\Big[\gamma^{\alpha}{S}_{u}^{bb^{\prime }}(x)\gamma^{\beta}S_{c}^{b^{\prime }b}(-x)\Big]
 \Big\}| 0 \rangle_\gamma,
\end{align} 
\begin{align}
\label{neweq1}
\Pi _{\mu \nu }^{\mathrm{QCD-Z_c^1}}(p,q)&=\frac{i}{2}
\int d^{4}xe^{ipx} \langle 0 | \Big\{ 
\mathrm{Tr}\Big[\gamma _{5}{S}_{c}^{aa^{\prime }}(x)\gamma _{5}S_{d}^{a^{\prime }a}(-x)\Big]
\mathrm{Tr}\Big[\gamma _{\mu }{S}_{u}^{bb^{\prime }}(x)\gamma _{\nu}S_{c}^{b^{\prime }b}(-x)\Big] \notag \\
&+\mathrm{Tr}\Big[ \gamma _{5 }{S}_{c}^{aa^{\prime}}(x)\gamma _{\nu}S_{d}^{a^{\prime }a}(-x)\Big]
\mathrm{Tr}\Big[ \gamma_{\mu }{S}_{u}^{bb^{\prime }}(x)\gamma _{5}S_{c}^{b^{\prime }b}(-x)\Big] \notag \\
&+\mathrm{Tr}\Big[\gamma _{\mu}{S}_{c}^{aa^{\prime }}(x)\gamma _{5 }S_{d}^{a^{\prime}a}(-x)\Big] 
\mathrm{Tr}\Big[ \gamma _{5}{S}_{u}^{bb^{\prime}}(x)\gamma _{\nu }S_{c}^{b^{\prime }b}(-x)\Big] \notag \\
&+\mathrm{Tr}\Big[\gamma _{\mu }{S}_{c}^{aa^{\prime }}(x)\gamma _{\nu }S_{d}^{a^{\prime}a}(-x)\Big] 
\mathrm{Tr}\Big[\gamma _{5}{S}_{u}^{bb^{\prime }}(x)\gamma_{5}S_{c}^{b^{\prime }b}(-x)\Big]
 \Big\}| 0 \rangle_\gamma,
\end{align}
\begin{align}
\label{neweq2}
\Pi _{\mu \nu }^{\mathrm{QCD-Z_{cs}}}(p,q)&=\frac{i}{2}
\int d^{4}xe^{ipx} \langle 0 | \Big\{ 
\mathrm{Tr}\Big[\gamma _{5}{S}_{c}^{aa^{\prime }}(x)\gamma _{5}S_{d}^{a^{\prime }a}(-x)\Big]
\mathrm{Tr}\Big[\gamma _{\mu }{S}_{s}^{bb^{\prime }}(x)\gamma _{\nu}S_{c}^{b^{\prime }b}(-x)\Big] \notag \\
&+\mathrm{Tr}\Big[ \gamma _{5 }{S}_{c}^{aa^{\prime}}(x)\gamma _{\nu}S_{d}^{a^{\prime }a}(-x)\Big]
\mathrm{Tr}\Big[ \gamma_{\mu }{S}_{s}^{bb^{\prime }}(x)\gamma _{5}S_{c}^{b^{\prime }b}(-x)\Big] \notag \\
&+\mathrm{Tr}\Big[\gamma _{\mu}{S}_{c}^{aa^{\prime }}(x)\gamma _{5 }S_{d}^{a^{\prime}a}(-x)\Big] 
\mathrm{Tr}\Big[ \gamma _{5}{S}_{s}^{bb^{\prime}}(x)\gamma _{\nu }S_{c}^{b^{\prime }b}(-x)\Big] \notag \\
&+\mathrm{Tr}\Big[\gamma _{\mu }{S}_{c}^{aa^{\prime }}(x)\gamma _{\nu }S_{d}^{a^{\prime}a}(-x)\Big] 
\mathrm{Tr}\Big[\gamma _{5}{S}_{s}^{bb^{\prime }}(x)\gamma_{5}S_{c}^{b^{\prime }b}(-x)\Big]
 \Big\}| 0 \rangle_\gamma,
\end{align}
\begin{align}
\label{neweq3}
\Pi _{\mu \nu }^{\mathrm{QCD-Z^{1}_{cs}}}(p,q)&=\frac{i}{2}
\int d^{4}xe^{ipx} \langle 0 | \Big\{ 
\mathrm{Tr}\Big[\gamma^{\alpha}{S}_{c}^{aa^{\prime }}(x)\gamma ^{\beta}S_{u}^{a^{\prime }a}(-x)\Big] 
\mathrm{Tr}\Big[\sigma_{\mu\alpha}\gamma _{5 }{S}_{s}^{bb^{\prime }}(x)\gamma _{5}\sigma_{\nu\beta}S_{c}^{b^{\prime }b}(-x)\Big] \notag \\
&-\mathrm{Tr}\Big[ \gamma^{\alpha}{S}_{c}^{aa^{\prime}}(x)\gamma _{5}\sigma_{\nu\beta}S_{u}^{a^{\prime }a}(-x)\Big]   
\mathrm{Tr}\Big[ \sigma_{\mu\alpha}\gamma_{5 }{S}_{s}^{bb^{\prime }}(x)\gamma^{\beta}S_{c}^{b^{\prime }b}(-x)\Big] \notag \\
&-\mathrm{Tr}\Big[\sigma_{\mu\alpha}\gamma _{5}{S}_{c}^{aa^{\prime }}(x)\gamma^{\beta }S_{u}^{a^{\prime}a}(-x)\Big]   
\mathrm{Tr}\Big[ \gamma^{\alpha}{S}_{s}^{bb^{\prime}}(x)\gamma_{5}\sigma_{\nu\beta}S_{c}^{b^{\prime }b}(-x)\Big] \notag \\
&+\mathrm{Tr}\Big[\sigma_{\mu\alpha}\gamma_{5 }{S}_{c}^{aa^{\prime }}(x)\gamma _{5}\sigma_{\nu\beta}S_{u}^{a^{\prime}a}(-x)\Big]  
\mathrm{Tr}\Big[\gamma^{\alpha}{S}_{s}^{bb^{\prime }}(x)\gamma^{\beta}S_{c}^{b^{\prime }b}(-x)\Big]
 \Big\}| 0 \rangle_\gamma,
\end{align}
where $S_{q(c)}(x)$ is propagator of the light q = u, d, s quarks and c-quark, and it is identified as 

\begin{align}
\label{edmn12}
S_{q}(x) &= 
\frac{1}{2 \pi^2 x^2}\Big( i \frac{{\xslash}}{x^{2}}-\frac{m_{q}}{2 } \Big)
- \frac{\langle \bar qq \rangle }{12} \Big(1-i\frac{m_{q} \xslash}{4}   \Big)
- \frac{\langle \bar q \sigma.G q \rangle }{192}x^2  \Big(1-i\frac{m_{q} \xslash}{6}   \Big)
-\frac {i g_s }{32 \pi^2 x^2} ~G^{\mu \nu} (x) \bigg[\rlap/{x}
\sigma_{\mu \nu} +  \sigma_{\mu \nu} \rlap/{x}
 \bigg],
\end{align}%
\begin{align}
\label{edmn13}
S_{c}(x)&=\frac{m_{c}^{2}}{4 \pi^{2}} \Bigg[ \frac{K_{1}\Big(m_{c}\sqrt{-x^{2}}\Big) }{\sqrt{-x^{2}}}
+i\frac{{\xslash}~K_{2}\Big( m_{c}\sqrt{-x^{2}}\Big)}
{(\sqrt{-x^{2}})^{2}}\Bigg]
-\frac{g_{s}m_{c}}{16\pi ^{2}} \int_0^1 dv\, G^{\mu \nu }(vx)\Bigg[ (\sigma _{\mu \nu }{\xslash}
  +{\xslash}\sigma _{\mu \nu })
  \frac{K_{1}\Big( m_{c}\sqrt{-x^{2}}\Big) }{\sqrt{-x^{2}}}
  \nonumber\\
  &
+2\sigma_{\mu \nu }K_{0}\Big( m_{c}\sqrt{-x^{2}}\Big)\Bigg].
\end{align}%
The correlation functions in Eqs. (\ref{neweq})-(\ref{neweq3}) includes various contributions: the photon can be emitted perturbatively and non-perturbatively. 
The next step is to calculate these contributions. These procedures are standard in the LCSR but quite lengthy. Therefore, we do not present these steps. The interested readers can find details of such calculations in Ref. \cite{Ozdem:2017exj}. With these calculations, the results of the QCD representation were obtained.
We can move on to calculate the results for hadronic representation.

The correlation function in Eq. ({\ref{edmn01}}) can also be re-written by inserting the a complete set of the $Z_{c(s)}$ tetraquark currents with the same quantum numbers into the Eq. (\ref{curr}) as follows:
 
\begin{align}
\label{edmn04}
\Pi_{\mu\nu}^{Had} (p,q) &= {\frac{\langle 0 \mid J_\mu (x) \mid
Z_{c(s)}(p) \rangle}{p^2 - m_{Z_{c(s)}}^2}}
\langle Z_{c(s)}(p) \mid Z_{c(s)}(p+q) \rangle_\gamma 
\frac{\langle Z_{c(s)}(p+q) \mid {J^\dagger}_\nu (0) \mid 0 \rangle}{(p+q)^2 - m_{Z_{c(s)}}^2} + \cdots,
\end{align}
where dots stands for the contributions coming from the higher states and continuum. 
%
 In the existence of the external electromagnetic background field, 
the matrix element $\langle Z_{c(s)}(p) \mid  Z_{c(s)}(p+q)\rangle_\gamma $ can be expressed associated with the Lorentz invariant form factors as follows~\cite{Brodsky:1992px}:

\begin{align}
\label{edmn06}
\langle Z_{c(s)}(p) \mid  Z_{c(s)}(p+q)\rangle_\gamma
 &= - \varepsilon^\tau (\varepsilon^{\theta})^\alpha
(\varepsilon^{\delta})^\beta \Big[ G_1(Q^2) 
(2p+q)_\tau ~g_{\alpha\beta} 
+
G_2(Q^2)
( g_{\tau\beta}~ q_\alpha -  g_{\tau\alpha}~ q_\beta)  \nonumber\\
& - \frac{G_3(Q^2)}{2 m_{Z_{c(s)}}^2}   (2p+q)_\tau ~q_\alpha q_\beta  \Big],
\end{align}
where $\varepsilon^\delta$ and $\varepsilon^{\theta}$ are the 
polarization vectors of the initial and final $Z_{c(s)}$
states and, q and $\varepsilon^\tau$ is the momentum and  polarization vector of the photon, respectively. 

The magnetic form factors $F_M(Q^2)$ can be written in the following way: 
\begin{align}
\label{edmn07}
&F_M(Q^2) = G_2(Q^2)\,,
\end{align}
where $Q^2=-q^2$.  At zero momentum transfer, i.e., $Q^2 = 0 $, the form
factors $F_M(0)$, is proportional to the
 magnetic moment $\mu_{Z_{c(s)}}$ in
the following way:
\begin{align}
\label{edmn08}
&\mu_{Z_{c(s)}} = \frac{ e}{2\, m_{Z_{c(s)}}} \,F_M(0).
\end{align}
Using Eqs. (\ref{edmn04})-(\ref{edmn06}), the correlation function takes the form,

\begin{align}
\label{edmn09}
 \Pi_{\mu\nu}^{Had}(p,q) &=   \frac{\lambda_{Z_{c(s)}}^2}{ [m_{Z_{c(s)}}^2 - (p+q)^2][m_{Z_{c(s)}}^2 - p^2]}
 \Big[ F_M (0) \Big(q_\mu \varepsilon_{\nu}
 - q_\nu \varepsilon_{\mu} +
\frac{\varepsilon.p}{m_{Z_{c(s)}}^2}  (p_\mu q_\nu - p_\nu q_\mu ) \Big)
\nonumber\\
&+\mbox{other independent form factors}\Big]\,.
\end{align}

The final step for analytical calculations will be to match the results obtained using two different representations. 
To do this, appropriate Lorentz structures are selected and Borel transform and continuum subtraction are applied to eliminate the contribution of higher states and continuum terms. Finally, we choose the structure $q_\mu \varepsilon_\nu$ for the MDMs and we get 
\begin{align}
 &\mu_{Z_{c}}\,\, \lambda_{Z_{c}}^2  = e^{\frac{m_{Z_{c}}^2}{M^2}} \,\, \Delta_1^{QCD},\\
 &\mu_{Z^1_{c}}\,\, \lambda_{Z^1_{c}}^2  = e^{\frac{m_{Z^1_{c}}^2}{M^2}} \,\, \Delta_2^{QCD},\\
  &\mu_{Z_{cs}}\,\, \lambda_{Z_{cs}}^2  =e^{\frac{m_{Z_{cs}}^2}{M^2}}\,\, \Delta_3^{QCD},\\
  &\mu_{Z^1_{cs}}\,\, \lambda_{Z^1_{cs}}^2  =e^{\frac{m_{Z^1_{cs}}^2}{M^2}}\,\, \Delta_4^{QCD}.
 \end{align}

The explicit expressions of the  $\Delta_i^{QCD}$ functions are presented in the appendix.

 \end{widetext}

\section{Numerical analysis} \label{secIII}

In this section, we numerically analyze the result of calculations for the MDMs.
We use $m_u=m_d=0$, $m_s =96^{+8.0}_{-4.0}\,\mbox{MeV}$,
$m_c = (1.275\pm 0.025)\,$GeV,  
$m_{Z_{c}}= 4022.9 \pm 0.8  \pm 2.7\, \mbox{MeV}$,
$m_{Z^1_{c}}= 4196^{+31+17}_{-29-13}\, \mbox{MeV}$,
$m_{Z_{cs}}= 4003.0\pm 6.0^{+4.0}_{-14.0}\, \mbox{MeV}$, $m_{Z^1_{cs}}= 4216.0 \pm 24.0^{+43}_{-30}~\mbox{MeV}$, 
$f_{3\gamma}=-0.0039~$GeV$^2$~\cite{Ball:2002ps},  
$\chi=-2.85 \pm 0.5~\mbox{GeV}^{-2}$~\cite{Rohrwild:2007yt}
$\langle \bar ss\rangle $= $0.8 \langle \bar uu\rangle$,
$\langle \bar uu\rangle $= 
$\langle \bar dd\rangle$=$(-0.24\pm0.01)^3\,$GeV$^3$  \cite{Ioffe:2005ym},   
$m_0^{2} = 0.8 \pm 0.1$~GeV$^2$, $\langle g_s^2G^2\rangle = 0.88~ $GeV$^4$~\cite{Nielsen:2009uh},
To proceed calculations, we need numerical value of the residues of the $Z_{c(s)}$ states. They are borrowed from Refs.~\cite{Chen:2021erj,Chen:2015ata}, which were obtained in the framework of the the two-point QCD sum rules.  The parameters used in the photon distribution amplitudes are given in Ref.~\cite{Ball:2002ps}.

Besides these input parameters used, the expressions for the MDMs contain two auxiliary parameters: continuum threshold $s_0$ and Borel mass parameter $M^2$. The standard prescription in LCSR is that physical observables should be weakly dependent on auxiliary parameters. In other words, we need to find the working regions of these auxiliary parameters, on which MDMs are independent.
To obtain the working regions of these helping parameters, the standard way of the LCSR approach such as, weak dependence of the results on the helping parameters, convergence of the OPE and pole dominance are considered. 
 Our numerical calculations leads to the conclusion
that these constraints are satisfied in the intervals shown below for the considered tetraquark states.
\begin{align*}
&18.6~\mbox{GeV}^2 \leq s_0 \leq 20.6~\mbox{GeV}^2~~ \mbox{for}~ Z_{c}~ \mbox{state},\\
&4.5~\mbox{GeV}^2 \leq M^2 \leq 6.5~\mbox{GeV}^2 ~~ \mbox{for} ~ Z_{c}~ \mbox{state},\\
\nonumber\\
&20.3~\mbox{GeV}^2 \leq s_0 \leq 22.1~\mbox{GeV}^2~~\mbox{for}~ Z^1_{c}~ \mbox{state},\\
&5.0~\mbox{GeV}^2 \leq M^2 \leq 7.0~\mbox{GeV}^2 ~~ \mbox{for} ~ Z^1_{c}~ \mbox{state},
\\
\nonumber\\
&18.3~\mbox{GeV}^2 \leq s_0 \leq 20.1~\mbox{GeV}^2~~\mbox{for}~ Z_{cs}~ \mbox{state},\\
&4.0~\mbox{GeV}^2 \leq M^2 \leq 6.0~\mbox{GeV}^2 ~~ \mbox{for} ~ Z_{cs}~ \mbox{state},
\end{align*}
\begin{align*}
&20.5~\mbox{GeV}^2 \leq s_0 \leq 22.5~\mbox{GeV}^2~~\mbox{for}~ Z^1_{cs}~ \mbox{state},\\
&5.0~\mbox{GeV}^2 \leq M^2 \leq 7.0~\mbox{GeV}^2 ~~ \mbox{for} ~ Z^1_{cs}~ \mbox{state}.
\end{align*}

Since we have numerical values of all input parameters, we can start numerical calculations. In Fig. 1, we plot the dependencies of the MDMs on $M^2$ at three values of $s_0$.   As is seen, the MDMs of $Z_{c(s)}$ states shows good stability with respect to $M^2$ in its working window. While MDMs shows some dependence on $s_0$, it remains inside the limits allowed by the LCSR and constitutes the major parts of the ambiguities. 
\begin{widetext}

\begin{figure}[htp]
\centering
 \includegraphics[width=0.455\textwidth]{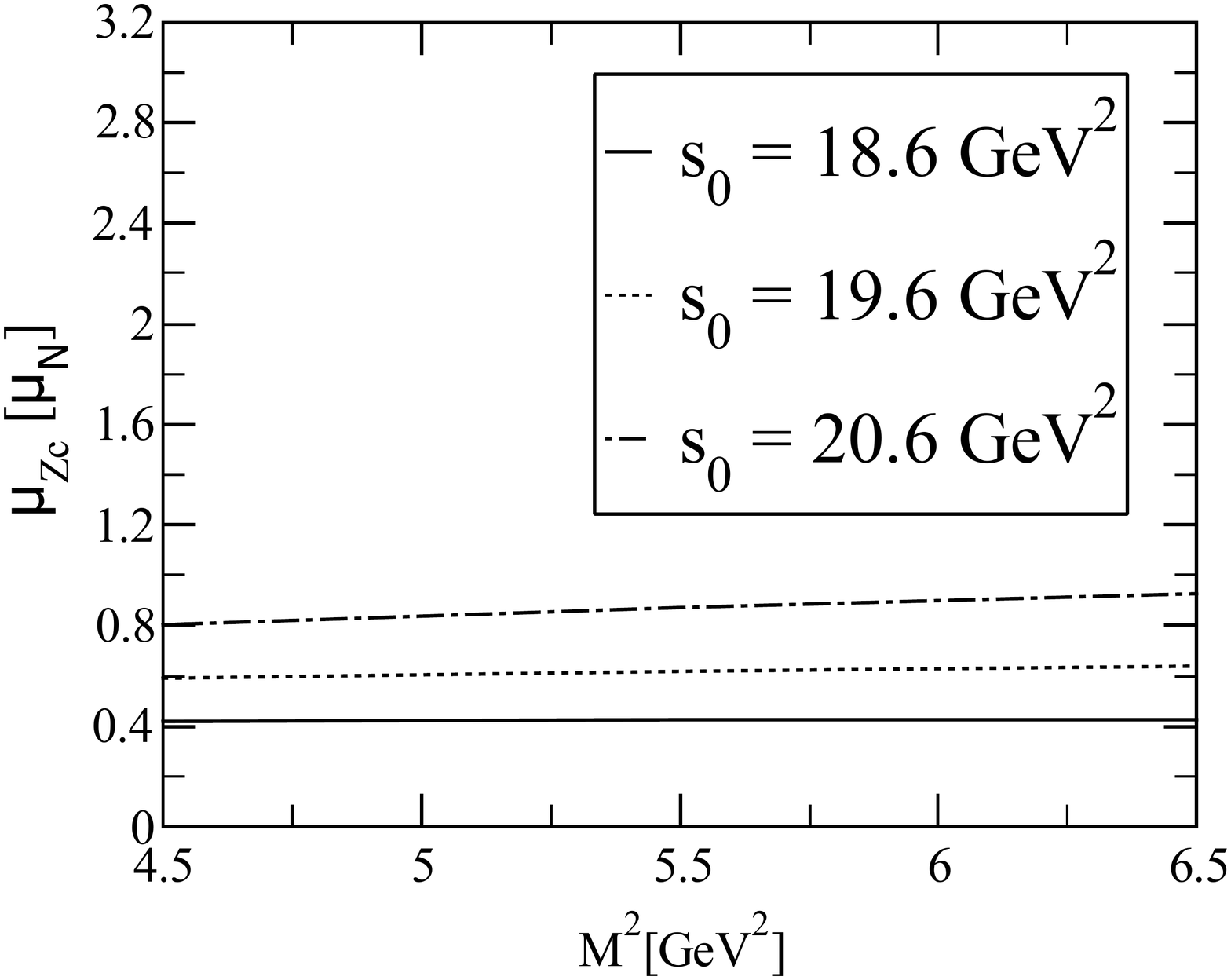}~~
\includegraphics[width=0.455\textwidth]{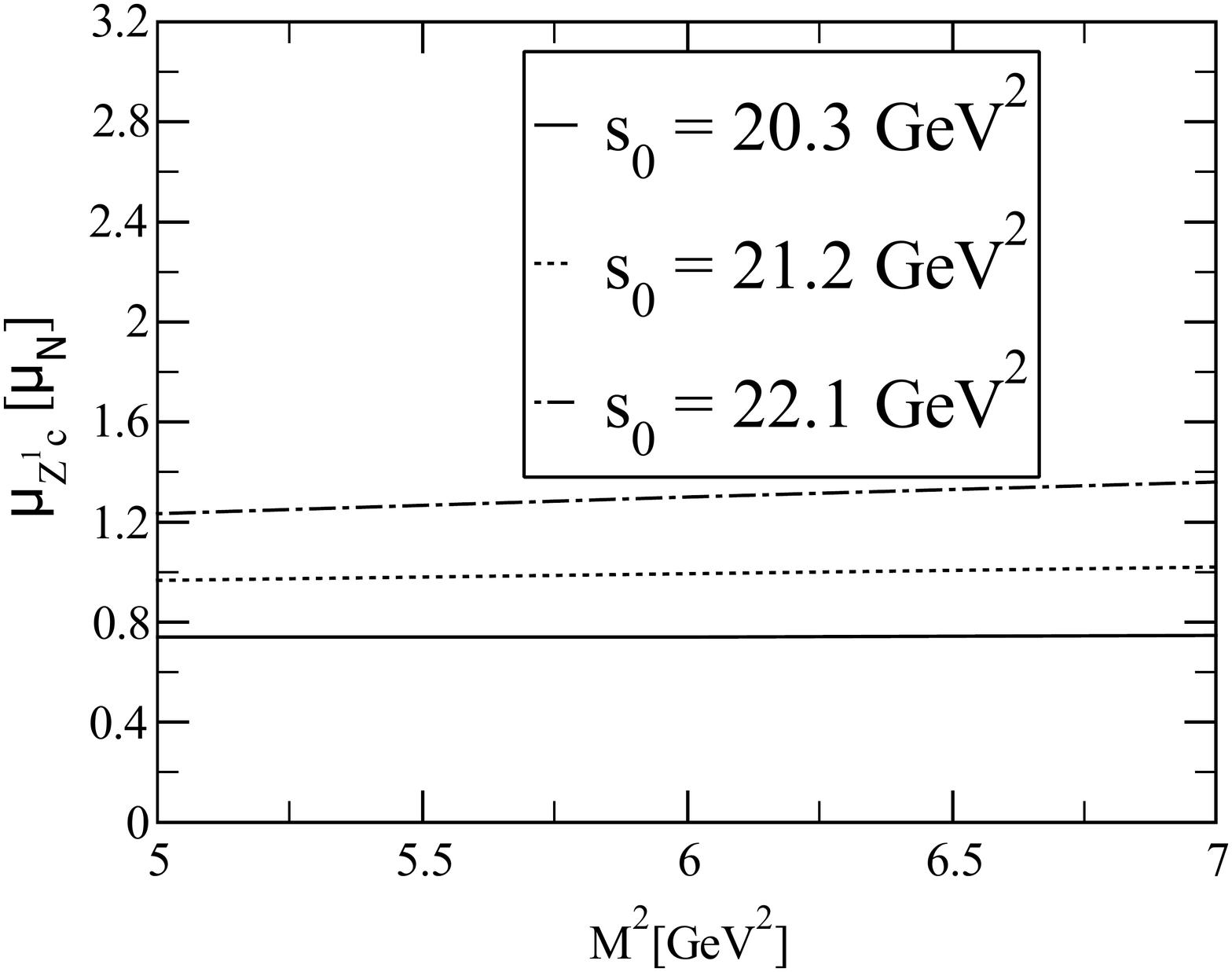}\\
  \vspace*{0.3cm}
\includegraphics[width=0.455\textwidth]{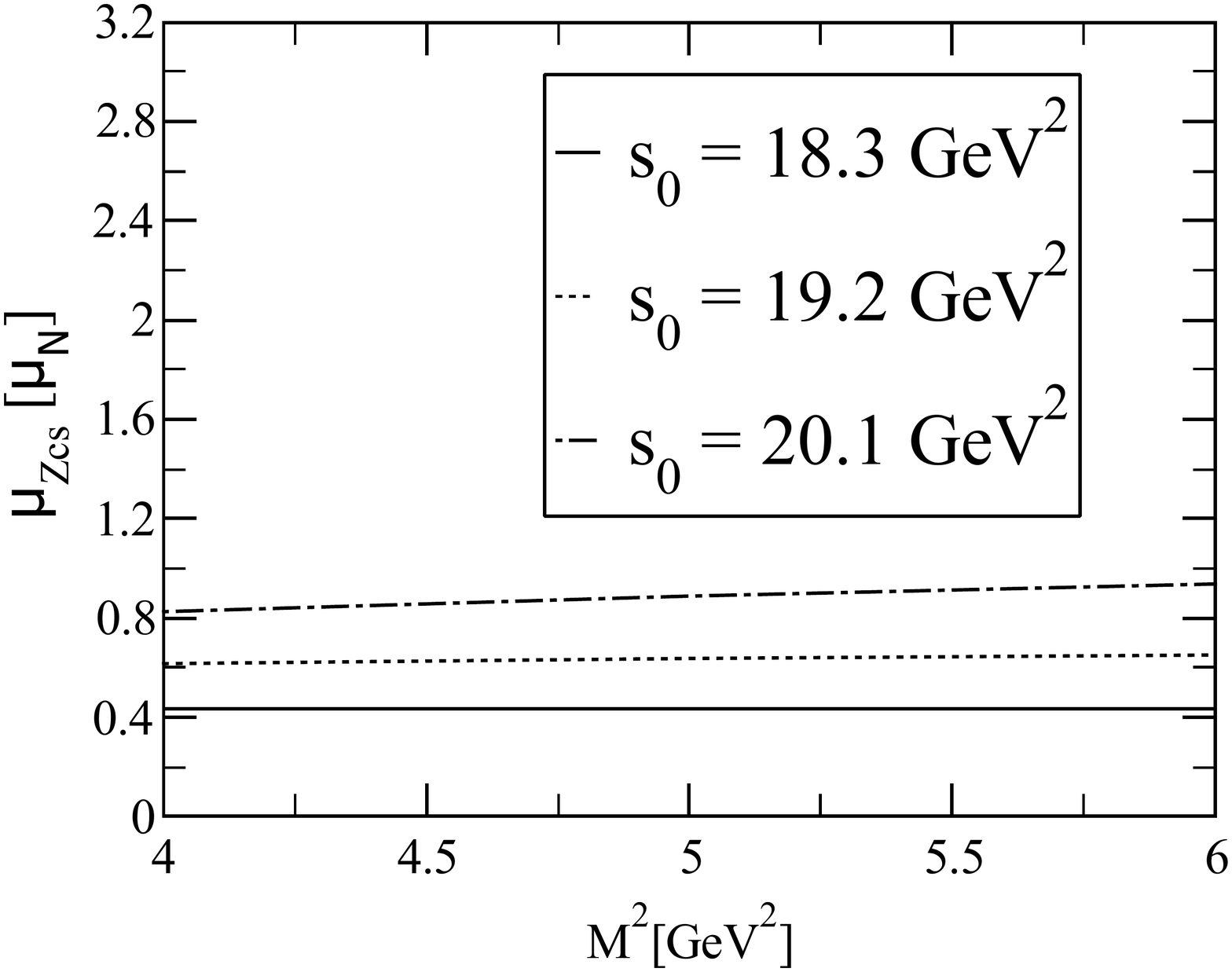}~~
\includegraphics[width=0.455\textwidth]{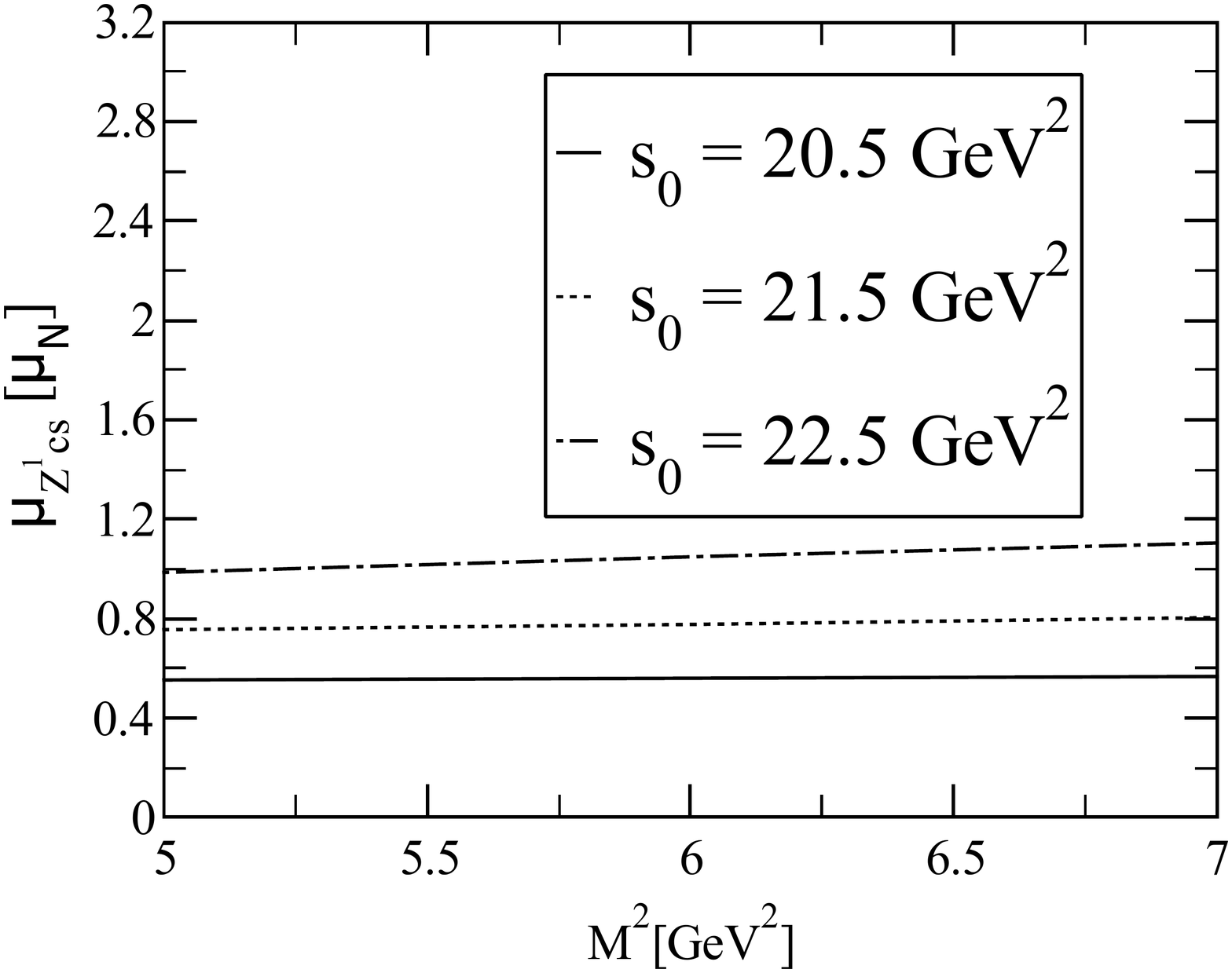}
 \caption{ The dependencies of MDMs of the $\mu_{Z_{c}}$, $\mu_{Z^1_{c}}$, $\mu_{Z_{cs}}$ and $\mu_{Z^1_{cs}}$ states; on $M^{2}$
 at three different values of $s_0$.}
  \end{figure}
  
  \end{widetext}
 
Our final results for the MDMs of charged $Z_{c(s)}$ states are 
\begin{align}
 &                \mu_{Z_{c}} = 0.66^{+0.27}_{-0.25}~\mu_{N}, ~~~~\\
&                 \mu_{Z^{1}_{c}}=1.03^{+0.32}_{-0.29}~\mu_{N},
\\
   &              \mu_{Z_{cs}}=0.73^{+0.28}_{-0.26}~\mu_{N}, ~~~~\\
 &                \mu_{Z^1_{cs}}=0.77^{+0.27}_{-0.25}~\mu_{N}.
\end{align}
%
The uncertainties in the results are due to $ s_0 $, $ M^2 $ and other input parameters. We should note that the main source of uncertainties is $s_0$.
It is seen that the results obtained for these states are large enough to be measured experimentally. 
Unfortunately, there are no experimental data and theoretical results that we can compare.

In order to obtain a more profound comprehension of the underlying quark-gluon dynamics, it is useful to consider the individual quark sector contributions to the MDMs. This can be achieved by dialing the corresponding charge factors $e_u$, $e_d$, $e_s$ and $e_c$. When it was done in case of $Z_c$ and $Z^1_c$ states we see that terms proportional the $e_u$ contributes about $\%67$ to the total MDMs, $e_d$ about $\%33$, and $e_c$ almost zero. In case of $Z_{cs}$ and $Z^1_{cs}$ states we observe that terms proportional the $e_u$ contributes about $\%69$ to the total MDMs, $e_s$ about $\%31$, and $e_c$ almost zero. A closer investigation shows that the smallness of the $e_c$ contribution is because of an almost exact cancellation of the terms involving the $e_c$. 
%

As we mentioned in the introduction, the $Z_{cs}(3985)$ and $Z_{cs}$ states have very different decay widths and can be considered as two different states, in spite of being very close in mass and also there is a possibility that $Z_{cs}(3985)$ and $Z_{cs}$ states are the identical, but their masses and decay widths may not be consistent because of coupled-channels effects or experimental resolution. In the literature, there are few attempts to have some assignments on the substructure of $Z_{cs}$ and $Z^1_{cs}$ states.
In Ref. \cite{Meng:2020ihj}, they showed that the $Z_{cs}(3985)$  state is the $U$-spin partner of $Z_{c}(3900)$ as a resonance within coupled-channel calculation in the $SU(3)_f$ symmetry and heavy quark spin symmetry. They were predicted a new the $Z_{cs}$ state with a mass $\sim$ 4130 MeV, which is the $U/V$-spin partner state of the charged $Z_c(4020)$ and heavy quark spin symmetry partner state of the $Z_{cs}(3985)$.
In Ref. \cite{ Ortega:2021enc}, structures of the $Z_c$ and $Z_{cs}(3985)$ states have been obtained using the chiral constituent quark model in a coupled-channels calculation. They obtained that the $Z_{cs}(3985)$ and $Z_{cs}(4000)$ are the same state. They also predicted a new state, the $SU(3)_f$ partner of the $Z_c(4020)$ state at $\sim$ 4110 MeV.
In Ref.~\cite{Wang:2020htx}, the mass spectrum of the $Z_{cs}(3985)$ state have been obtained in the framework of the chiral perturbation theory.  They predicted a new state with a mass around the $\sim$4125 MeV, which can be strange partner of $Z_c(4020)$ state.
In Ref. \cite{ Chen:2021erj}, mass, decay constants and production rates of the $Z_{c(s)}$ states were investigated in molecular picture within the QCD sum rules. The obtained results supported that the the $Z_{cs}(3985)$, $Z_{cs}(4000)$ and  $Z_{cs}(4220)$ were strange partners of the $X(3872)$, $Z_c(3900)$ and $Z_c(4020)$, respectively.
In Ref. \cite{Meng:2021rdg}, they studied the properties of the $Z_{cs}(3985)$ and $Z_{cs}(4000)$ states in the molecular picture via solvable non-relativistic effective field theory. They found that $Z_{cs}(3985)$ and $Z_{cs}(4000)$ states are different states and $Z_{cs}(4220)$ state is heavy quark spin symmetry partner state of the $Z_{cs}(4000)$. Also, they predicted that  $Z_{cs}(4000)$ and $Z_{cs}(4220)$ states are the $SU(3)_f$ partners of the $Z_{c}(3900)$ and $Z_{c}(4020)$ states, respectively.
In Ref. \cite{Ozdem:2021yvo}, the MDM of $Z_{cs}(3985)$ state have been acquired in the framework of LCSR  by using the molecular and compact diquark-antidiquark type interpolating currents. The obtained results are given as $\mu_{Z_{cs}}^{Di} =0.60^{+0.26}_{-0.21}~\mu_N$ and  $\mu_{Z_{cs}}^{Mol} =0.52^{+0.19}_{-0.17} ~\mu_N$ for diquark-antidiquark and molecular pictures, respectively. 
When we compare our results, it can be seen that the difference between the central values is large enough to say that these two states are different states. Therefore, our MDM results are support that $Z_{cs}(3985)$ and $Z_{cs}$ are different states.
We also mentioned that the $Z^1_{cs}$ state can be interpreted as the strange partner of the $Z_c$ state. 
We predict that if the $Z_{c}$ and $Z^1_{cs}$ states are $SU(3)_f$ partners, the difference between their MDMs should be around $\%15$.
If this difference is more than $\%15$, these states are not $SU(3)_f$ partners.
As can be seen from studies of the $Z_c$ and $Z_{cs}$ states, the results achieved using different approaches give rise to different estimations. The properties of these states should be tested in more precise experimental data in the future and we need more experimental and theoretical studies on the different decay channels of $Z_{c(s)}$ to figure out their inner configurations.

Our final remark is that the MDMs of the neutral $Z_{cs}$ and $Z^1_{cs}$ states have also been extracted. 
The results are presented as
\begin{align}
    &       \mu_{Z^{0}_{cs}}=(3.0^{+0.9}_{-0.8})\times 10^{-2}~\mu_{N},\\
    \nonumber\\
   &       \mu_{Z^{1,0}_{cs}}=(3.6^{+0.9}_{-0.9})\times 10^{-2}~\mu_{N}.
\end{align}
 %
The MDMs of the  neutral $Z_{cs}$ and $Z^1_{cs}$ states are obtained nonzero but small values. The nonzero values of the MDMs of  neutral $Z_{cs}$ and $Z^1_{cs}$ states are due to the $SU(3)_f$ symmetry breaking.

\section{Discussion and concluding remarks}\label{secIV}

To investigate the hidden-charm tetraquark states with and without strangeness, we have acquired the magnetic dipole moments for the molecular picture with $J^P = 1^{+}$ quantum numbers within light-cone QCD sum rules. For this aim, we use the relevant molecular interpolating currents and the photon distribution amplitudes.
The electromagnetic form factors of particles encode the spatial distributions of charge and magnetization in the particles and ensure significant knowledge about the quark organizations of the particles and the underlying dynamics.
%
The size of the results on $Z_{c}(4020)^+$, $Z_{c}(4200)^+$, $Z_{cs}(4000)^{+}$ and  $Z_{cs}(4220)^{+}$ states indicates that the magnetic dipole moments of these states are accessible in the experiment. 
The magnetic dipole moment of the neutral $Z_{cs}(4000)$ and  $Z_{cs}(4220)$ states are also nonzero, however their values are small.
The direct measurement of the magnetic dipole moments of exotic states are not likely in the near future.  For this reason, any indirect predictions of the magnetic dipole moments of the exotic states would be very helpful.
Any experimental measurements of the magnetic dipole moments of the hidden-charm tetraquark states and comparison of the acquired results with the estimations of the this work may ensure as useful information on the substructure of the these tetraquark states as well as the non-perturbative behaviors of the strong interaction at the low-energy region.

\section{Acknowledgements}

We are grateful to A. Ozpineci for useful discussions, comments and remarks.

\begin{widetext}
\section*{Appendix: Explicit forms of the  \texorpdfstring{$\Delta_i^{QCD}$}{} functions } 
In this appendix, we give the explicit expressions for the  $\Delta_i^{QCD}$ functions: 
\begin{align}
\Delta_1^{QCD}&=\frac {1}{47185920 \pi^5}\Bigg\{4 e_c m_c m_q \Bigg (20  P_ 1 \Big (7 I[0, 3, 1, 0] - 20 I[0, 3, 1, 1] + 19 I[0, 3, 1, 2] - 
          6 I[0, 3, 1, 3]- 14 I[0, 3, 2, 0]\nonumber\\
          & + 26 I[0, 3, 2, 1] - 
          12 I[0, 3, 2, 2] + 7 I[0, 3, 3, 0] - 6 I[0, 3, 3, 1] + 
          18 I[1, 2, 1, 1] - 36 I[1, 2, 1, 2]+ 18 I[1, 2, 1, 3]\nonumber\\
          & - 
          36 I[1, 2, 2, 1] + 36 I[1, 2, 2, 2] + 
          18 I[1, 2, 3, 1]\Big) + 
       27 \Big (6 I[0, 5, 1, 1] - 18 I[0, 5, 1, 2] + 
           18 I[0, 5, 1, 3]- 6 I[0, 5, 1, 4] \nonumber\\
           & - 16 I[0, 5, 2, 1] + 
           31 I[0, 5, 2, 2] - 14 I[0, 5, 2, 3] - I[0, 5, 2, 4] + 
           14 I[0, 5, 3, 1] - 12 I[0, 5, 3, 2] - 2 I[0, 5, 3, 3] \nonumber\\
           &- 
           4 I[0, 5, 4, 1] - I[0, 5, 4, 2] + 5 I[1, 4, 2, 2] - 
           10 I[1, 4, 2, 3] + 5 I[1, 4, 2, 4] - 10 I[1, 4, 3, 2] + 
           10 I[1, 4, 3, 3] \nonumber\\
           &+ 5 I[1, 4, 4, 2]\Big)\Bigg)
           \nonumber\\
           &- 
    e_d \Bigg (-15 P_ 1 \Big (I[0, 4, 2, 0] - 3 I[0, 4, 2, 1] + 
          3 I[0, 4, 2, 2] - I[0, 4, 2, 3] - 3 I[0, 4, 3, 0] + 
          6 I[0, 4, 3, 1] - 3 I[0, 4, 3, 2] \nonumber\\
          &+ 3 I[0, 4, 4, 0] - 
          3 I[0, 4, 4, 1] - I[0, 4, 5, 0] + 12 I[1, 3, 2, 1] - 
          12 I[1, 3, 2, 2] + 4 I[1, 3, 2, 3] - 24 I[1, 3, 3, 1]\nonumber\\
          &+ 
          12 I[1, 3, 3, 2] + 12 I[1, 3, 4, 1]\Big) + 
       6480  \pi^2 m_c P_ 2 \Big (I[0, 4, 2, 0] - 3 I[0, 4, 2, 1] + 
          3 I[0, 4, 2, 2] - I[0, 4, 2, 3] \nonumber\\
          &- 3 I[0, 4, 3, 0] + 
          6 I[0, 4, 3, 1] - 3 I[0, 4, 3, 2] + 3 I[0, 4, 4, 0] - 
          3 I[0, 4, 4, 1] - I[0, 4, 5, 0] + 12 I[1, 3, 2, 1] \nonumber\\
          &- 
          12 I[1, 3, 2, 2] + 4 I[1, 3, 2, 3] - 24 I[1, 3, 3, 1] + 
          12 I[1, 3, 3, 2] + 12 I[1, 3, 4, 1]\Big) + 
       8 m_c^2 \Big (20 P_ 1 \big (I[0, 3, 1, 0] \nonumber\\
       &- 3 I[0, 3, 1, 1] + 
             3 I[0, 3, 1, 2] - I[0, 3, 1, 3] - 2 I[0, 3, 2, 0] + 
             4 I[0, 3, 2, 1] - 2 I[0, 3, 2, 2] + I[0, 3, 3, 0] \nonumber\\
             &- 
             I[0, 3, 3, 1] + 3 I[1, 2, 1, 1] - 6 I[1, 2, 1, 2] + 
             3 I[1, 2, 1, 3] - 6 I[1, 2, 2, 1] + 6 I[1, 2, 2, 2] + 
             3 I[1, 2, 3, 1]) \nonumber\\
             &+ 
          27 (2 I[0, 5, 2, 1] - 5 I[0, 5, 2, 2] + 4 I[0, 5, 2, 3] - 
              I[0, 5, 2, 4] - 4 I[0, 5, 3, 1] + 6 I[0, 5, 3, 2] - 
              2 I[0, 5, 3, 3] \nonumber\\
              &+ 2 I[0, 5, 4, 1] - I[0, 5, 4, 2] + 
              5 I[1, 4, 2, 2] - 10 I[1, 4, 2, 3] + 5 I[1, 4, 2, 4] - 
              10 I[1, 4, 3, 2] + 10 I[1, 4, 3, 3] \nonumber\\
              &+ 
              5 I[1, 4, 4, 2]\big)\Big) + 
       243 \big (I[0, 6, 3, 0] - 4 I[0, 6, 3, 1] + 6 I[0, 6, 3, 2] - 
           4 I[0, 6, 3, 3] + I[0, 6, 3, 4] - 3 I[0, 6, 4, 0] \nonumber\\
           &+ 
           9 I[0, 6, 4, 1] - 9 I[0, 6, 4, 2] + 3 I[0, 6, 4, 3] + 
           3 I[0, 6, 5, 0] - 6 I[0, 6, 5, 1] + 3 I[0, 6, 5, 2] - 
           I[0, 6, 6, 0] + I[0, 6, 6, 1] \nonumber\\
           &+ 6 I[1, 5, 3, 1] - 
           18 I[1, 5, 3, 2] + 18 I[1, 5, 3, 3] - 6 I[1, 5, 3, 4] - 
           18 I[1, 5, 4, 1] + 36 I[1, 5, 4, 2] - 18 I[1, 5, 4, 3] \nonumber\\
           &+ 
           18 I[1, 5, 5, 1] - 18 I[1, 5, 5, 2] - 
           6 I[1, 5, 6, 1]\big)\Bigg) \nonumber\\
           &+ 
    e_u \Bigg (-15 P_ 1 \Big (I[0, 4, 2, 0] - 3 I[0, 4, 2, 1] + 
           3 I[0, 4, 2, 2] - I[0, 4, 2, 3] - 3 I[0, 4, 3, 0] + 
           6 I[0, 4, 3, 1] - 3 I[0, 4, 3, 2] \nonumber\\
           &+ 3 I[0, 4, 4, 0] - 
           3 I[0, 4, 4, 1] - I[0, 4, 5, 0] + 12 I[1, 3, 2, 1] - 
           12 I[1, 3, 2, 2] + 4 I[1, 3, 2, 3] - 24 I[1, 3, 3, 1] \nonumber\\
              &+ 
           12 I[1, 3, 3, 2] + 12 I[1, 3, 4, 1]\Big) + 
        8 m_c^2 \Big (-720 \pi^2 m_d P_ 2  \big (I[0, 3, 2, 0] - 
              2 I[0, 3, 2, 1] + I[0, 3, 2, 2] - 2 I[0, 3, 3, 0] \nonumber\\
              &+ 
              2 I[0, 3, 3, 1] + I[0, 3, 4, 0] + 6 I[1, 2, 2, 1] - 
              3 I[1, 2, 2, 2] - 6 I[1, 2, 3, 1]\big) + 
           20 P_ 1 \Big (I[0, 3, 1, 0]- 3 I[0, 3, 1, 1] \nonumber\\
           & + 
              3 I[0, 3, 1, 2] - I[0, 3, 1, 3] - 2 I[0, 3, 2, 0] + 
              4 I[0, 3, 2, 1] - 2 I[0, 3, 2, 2] + I[0, 3, 3, 0]- 
              I[0, 3, 3, 1] + 3 I[1, 2, 1, 1] \nonumber\\
              & - 6 I[1, 2, 1, 2] + 
              3 I[1, 2, 1, 3] - 6 I[1, 2, 2, 1] + 6 I[1, 2, 2, 2] + 
              3 I[1, 2, 3, 1]\Big)+27 \big (2 I[0, 5, 2, 1] - 5 I[0, 5, 2, 2] \nonumber\\
              &+ 
               4 I[0, 5, 2, 3] - I[0, 5, 2, 4] - 4 I[0, 5, 3, 1] + 
               6 I[0, 5, 3, 2] - 2 I[0, 5, 3, 3] + 2 I[0, 5, 4, 1] - 
               I[0, 5, 4, 2] \nonumber\\
               &+ 5 I[1, 4, 2, 2] - 10 I[1, 4, 2, 3] + 
               5 I[1, 4, 2, 4] - 10 I[1, 4, 3, 2] + 
               10 I[1, 4, 3, 3] + 5 I[1, 4, 4, 2]\big)\Big)\nonumber\\
               &+ 
        324 m_c \Big (20 \pi^2 P_ 2  \big (I[0, 4, 2, 0] - 
              3 I[0, 4, 2, 1] + 3 I[0, 4, 2, 2] - I[0, 4, 2, 3] - 
              3 I[0, 4, 3, 0] + 6 I[0, 4, 3, 1] \nonumber\\
              &     - 3 I[0, 4, 3, 2] + 
              3 I[0, 4, 4, 0] - 3 I[0, 4, 4, 1] - I[0, 4, 5, 0] + 
              12 I[1, 3, 2, 1] - 12 I[1, 3, 2, 2] + 4 I[1, 3, 2, 3]\nonumber
           \end{align}
          \begin{align} 
          &             - 
              24 I[1, 3, 3, 1] + 12 I[1, 3, 3, 2] + 
              12 I[1, 3, 4, 1]\big) + 
           3 m_q \big (I[0, 5, 2, 0] - 4 I[0, 5, 2, 1] + 
               6 I[0, 5, 2, 2] - 4 I[0, 5, 2, 3] \nonumber\\
               &+ I[0, 5, 2, 4] - 
               3 I[0, 5, 3, 0] + 9 I[0, 5, 3, 1] - 9 I[0, 5, 3, 2] + 
               3 I[0, 5, 3, 3] + 3 I[0, 5, 4, 0] - 6 I[0, 5, 4, 1]+ 
               3 I[0, 5, 4, 2]\nonumber\\
               & - I[0, 5, 5, 0] + I[0, 5, 5, 1] + 
               5 I[1, 4, 2, 1] - 15 I[1, 4, 2, 2] + 
               15 I[1, 4, 2, 3] - 5 I[1, 4, 2, 4] - 
               15 I[1, 4, 3, 1] \nonumber\\
               &+ 30 I[1, 4, 3, 2] - 
               15 I[1, 4, 3, 3] + 15 I[1, 4, 4, 1] - 
               15 I[1, 4, 4, 2] - 5 I[1, 4, 5, 1]\big)\Big) + 
        243 \big (I[0, 6, 3, 0] \nonumber\\
        &- 4 I[0, 6, 3, 1] + 6 I[0, 6, 3, 2] - 
            4 I[0, 6, 3, 3] + I[0, 6, 3, 4] - 3 I[0, 6, 4, 0] + 
            9 I[0, 6, 4, 1] - 9 I[0, 6, 4, 2] \nonumber\\
            &+ 3 I[0, 6, 4, 3] + 
            3 I[0, 6, 5, 0] - 6 I[0, 6, 5, 1] + 3 I[0, 6, 5, 2] - 
            I[0, 6, 6, 0] + I[0, 6, 6, 1] + 6 I[1, 5, 3, 1] \nonumber\\
            &- 
            18 I[1, 5, 3, 2] + 18 I[1, 5, 3, 3] - 6 I[1, 5, 3, 4] - 
            18 I[1, 5, 4, 1] + 36 I[1, 5, 4, 2] - 18 I[1, 5, 4, 3] + 
            18 I[1, 5, 5, 1] \nonumber\\
            &- 18 I[1, 5, 5, 2] - 
            6 I[1, 5, 6, 1]\big)\Bigg)+20\,\pi^2\,f_{3\gamma} P_2 m_q 
 \Big(4\,(32e_d+41e_u)\,I[0,3,4,0]+3\,(9e_d+20e_u)\,I[0,3,5,0]\Big)I_2[\mathcal{V}]\nonumber\\
 &+3 m_0^2(e_d+14e_u)\,I_4[\tilde S]\, I[0,5,4,0]\Bigg\},
\end{align}

\begin{align}
 \Delta_2^{QCD}&=\frac {1} {10485760 \pi^5}\Bigg\{160 \pi^2 e_c m_c P_ 2  \Bigg (-3 I[
         0, 4, 1, 0] + 9 I[0, 4, 1, 1] - 9 I[0, 4, 1, 2] + 
       3 I[0, 4, 1, 3]+ 10 I[0, 4, 2, 0] \nonumber\\
       & - 21 I[0, 4, 2, 1] + 
       12 I[0, 4, 2, 2] - I[0, 4, 2, 3] - 11 I[0, 4, 3, 0] + 
       13 I[0, 4, 3, 1] - 2 I[0, 4, 3, 2]+ 4 I[0, 4, 4, 0] \nonumber\\
       & - 
       I[0, 4, 4, 1] + 
       m_ 0^2 (I[0, 3, 2, 0] - 2 I[0, 3, 2, 1] + I[0, 3, 2, 2] - 
          2 I[0, 3, 3, 0] + 2 I[0, 3, 3, 1]\nonumber\\
          &+ I[0, 3, 4, 0] + 
          6 I[1, 2, 2, 1] - 3 I[1, 2, 2, 2] - 6 I[1, 2, 3, 1]) + 
       4 I[1, 3, 2, 1] - 8 I[1, 3, 2, 2] + 4 I[1, 3, 2, 3] \nonumber\\
       & - 
       8 I[1, 3, 3, 1] + 8 I[1, 3, 3, 2]+ 4 I[1, 3, 4, 1]\Bigg) \nonumber\\
       &+ 
    e_u \Bigg (5 P_ 1 (I[0, 4, 2, 0] - 3 I[0, 4, 2, 1] + 
          3 I[0, 4, 2, 2] - I[0, 4, 2, 3] - 3 I[0, 4, 3, 0] + 
          6 I[0, 4, 3, 1] - 3 I[0, 4, 3, 2] \nonumber\\
          &+ 3 I[0, 4, 4, 0] - 
          3 I[0, 4, 4, 1] - I[0, 4, 5, 0] + 12 I[1, 3, 2, 1] - 
          12 I[1, 3, 2, 2] + 4 I[1, 3, 2, 3] - 24 I[1, 3, 3, 1]\nonumber\\
          &+ 
          12 I[1, 3, 3, 2] + 12 I[1, 3, 4, 1]) - 
       240 \pi^2 m_c P_ 2 \Big (I[0, 4, 2, 0] - 3 I[0, 4, 2, 1] + 
          3 I[0, 4, 2, 2] - I[0, 4, 2, 3] \nonumber\\
          &- 3 I[0, 4, 3, 0] + 
          6 I[0, 4, 3, 1] - 3 I[0, 4, 3, 2] + 3 I[0, 4, 4, 0] - 
          3 I[0, 4, 4, 1] - I[0, 4, 5, 0] + 12 I[1, 3, 2, 1]\nonumber\\
          &- 
          12 I[1, 3, 2, 2] + 4 I[1, 3, 2, 3] - 24 I[1, 3, 3, 1] + 
          12 I[1, 3, 3, 2] + 12 I[1, 3, 4, 1]\Big) - 
       9 \Big (I[0, 6, 3, 0] - 4 I[0, 6, 3, 1] \nonumber\\
       &+ 6 I[0, 6, 3, 2] - 
           4 I[0, 6, 3, 3] + I[0, 6, 3, 4] - 3 I[0, 6, 4, 0] + 
           9 I[0, 6, 4, 1] - 9 I[0, 6, 4, 2] + 3 I[0, 6, 4, 3] \nonumber\\
           &+ 
           3 I[0, 6, 5, 0] - 6 I[0, 6, 5, 1] + 3 I[0, 6, 5, 2] - 
           I[0, 6, 6, 0] + I[0, 6, 6, 1] + 6 I[1, 5, 3, 1] - 
           18 I[1, 5, 3, 2] \nonumber\\
           &+ 18 I[1, 5, 3, 3] - 6 I[1, 5, 3, 4] - 
           18 I[1, 5, 4, 1] + 36 I[1, 5, 4, 2] - 18 I[1, 5, 4, 3] + 
           18 I[1, 5, 5, 1] - 18 I[1, 5, 5, 2]\nonumber\\
           &- 
           6 I[1, 5, 6, 1]\Big)\Bigg) \nonumber\\
           &+ 
    e_d \Bigg (-5 P_ 1 \Big(I[0, 4, 2, 0] - 3 I[0, 4, 2, 1] + 
           3 I[0, 4, 2, 2] - I[0, 4, 2, 3] - 3 I[0, 4, 3, 0] + 
           6 I[0, 4, 3, 1] - 3 I[0, 4, 3, 2] \nonumber\\
           &+ 3 I[0, 4, 4, 0] - 
           3 I[0, 4, 4, 1] - I[0, 4, 5, 0] + 12 I[1, 3, 2, 1] - 
           12 I[1, 3, 2, 2] + 4 I[1, 3, 2, 3] - 24 I[1, 3, 3, 1] \nonumber\\
           &+ 
           12 I[1, 3, 3, 2] + 12 I[1, 3, 4, 1]\Big) + 
        240 \pi^2 m_c P_ 2 \Big (I[0, 4, 2, 0] - 3 I[0, 4, 2, 1] + 
           3 I[0, 4, 2, 2] - I[0, 4, 2, 3] \nonumber\\
           &- 3 I[0, 4, 3, 0] + 
           6 I[0, 4, 3, 1] - 3 I[0, 4, 3, 2] + 3 I[0, 4, 4, 0] - 
           3 I[0, 4, 4, 1] - I[0, 4, 5, 0] + 12 I[1, 3, 2, 1] \nonumber\\
           &- 
           12 I[1, 3, 2, 2] + 4 I[1, 3, 2, 3] - 24 I[1, 3, 3, 1] + 
           12 I[1, 3, 3, 2] + 12 I[1, 3, 4, 1]\Big) + 
        9 \Big (I[0, 6, 3, 0] - 4 I[0, 6, 3, 1] \nonumber\\
        &+ 6 I[0, 6, 3, 2] - 
            4 I[0, 6, 3, 3] + I[0, 6, 3, 4] - 3 I[0, 6, 4, 0] + 
            9 I[0, 6, 4, 1] - 9 I[0, 6, 4, 2] + 3 I[0, 6, 4, 3] \nonumber\\
            &+ 
            3 I[0, 6, 5, 0] - 6 I[0, 6, 5, 1] + 3 I[0, 6, 5, 2] - 
            I[0, 6, 6, 0] + I[0, 6, 6, 1] + 6 I[1, 5, 3, 1] - 
            18 I[1, 5, 3, 2] \nonumber\\
            &+ 18 I[1, 5, 3, 3] - 6 I[1, 5, 3, 4] - 
            18 I[1, 5, 4, 1] + 36 I[1, 5, 4, 2] - 18 I[1, 5, 4, 3] + 
            18 I[1, 5, 5, 1] - 18 I[1, 5, 5, 2] \nonumber
            \end{align}
        \begin{align}
            &- 
            6 I[1, 5, 6, 1]\Big)\Bigg)
            + P_1 \Big[ (6e_u-7e_d)
            \Big (I[0, 4, 3,1] - 3 I[0, 4, 4,0] + 3 I[0, 4, 5,2] - I[0, 4, 6,2] +
    4 I[1, 3, 3,1]  \nonumber\\
    &- 4 I[1, 3, 6,2] +  2 I[2, 2, 3,3] - 6 I[2, 2, 4,2] + 6 I[2, 2, 5,0] -
    2 I[2, 2, 6,4]\Big)\, \mathbb A[u_ 0]
    -5 (11e_u+13e_d) \bigg \{\Big (I[0, 4, 2,0] \nonumber\\
    &- I[0, 4, 5,1] + 2 I[1, 3, 2,1] - 6 I[1, 3, 3,2] + 6 I[1, 3, 4,2] \Big) \Big (I_ 2[\mathcal S] -
      2 I_ 2[\mathcal T_ 2] - I_ 2[\mathcal T_ 3] +
      I_ 2[\mathcal T_ 4]\Big)+ \Big (I[0, 4, 2,3]
    \nonumber\\
    &- 3 I[0, 4, 3,0] + 3 I[0, 4, 4,1] -  I[0, 4, 5,1]\Big)\Big (
      I_ 2[\mathcal T_ 1] + I_ 2[\mathcal {\tilde S}]\Big)\bigg\}
      +15 (5e_u-11e_d) \Big (I[0, 4, 3,1] - 3 I[0, 4, 4,1]  \nonumber\\
       &+     2 I[1, 3, 3,2]- 6 I[1, 3, 4,1] + 6 I[1, 3, 5,1] - 2 I[1, 3, 6,1]\Big)\, I_3[h_\gamma] -
 3\, \chi \, \Big (2 I[0, 5, 3,2] - 6 I[0, 5, 4,2] + 6 I[0, 5, 5,2]  \nonumber\\
  & -     2 I[0, 5, 6,2]+ 5 I[1, 4, 3,2]- 15 I[1, 4, 4,1] + 15 I[1, 4, 5,0] -
    5 I[1, 4, 6,0]\Big) \varphi_\gamma[u_ 0]\Big]\Bigg\},
\end{align}

\begin{align}
 \Delta_3^{QCD}&=\frac {1} {283115520 \pi^5}\Bigg\{16 e_Q m_c \Bigg (-40 m_c m_s P_ 1 \
P_ 3 \pi^2 (I[0, 1, 1, 0] - I[0, 1, 1, 1] - I[0, 1, 2, 0]) \nonumber\\&+ 
       360 \pi^2  (P_ 2 - P_ 3) \Big (-3 I[0, 4, 1, 0] + 
          9 I I[0, 4, 1, 1] - 9 I[0, 4, 1, 2] + 3 I[0, 4, 1, 3] + 
          10 I[0, 4, 2, 0] - 21 I[0, 4, 2, 1] \nonumber\\
          &+ 12 I[0, 4, 2, 2] - 
          I[0, 4, 2, 3] - 11 I[0, 4, 3, 0] + 13 I[0, 4, 3, 1] - 
          2 I[0, 4, 3, 2] + 4 I[0, 4, 4, 0] - I[0, 4, 4, 1]\nonumber\\
          &+ 
          m_ 0^2 \big(I[0, 3, 2, 0] - 2 I[0, 3, 2, 1] + I[0, 3, 2, 2] - 
             2 I[0, 3, 3, 0] + 2 I[0, 3, 3, 1] + I[0, 3, 4, 0] + 
             6 I[1, 2, 2, 1] \nonumber\\
             &- 3 I[1, 2, 2, 2] - 6 I[1, 2, 3, 1]\big) + 
          4 I[1, 3, 2, 1] - 8 I[1, 3, 2, 2] + 4 I[1, 3, 2, 3] - 
          8 I[1, 3, 3, 1] + 8 I[1, 3, 3, 2] \nonumber\\
          &+ 4 I[1, 3, 4, 1]\Big) + 
       m_s \Big (35 P_ 1 (I[0, 3, 1, 0] - 2 I[0, 3, 1, 1] + 
              I[0, 3, 1, 2] - 2 I[0, 3, 2, 0] + 2 I[0, 3, 2, 1] + 
              I[0, 3, 3, 0]) \nonumber\\
              &- 
           27 \big (6 I[0, 5, 1, 1] - 18 I[0, 5, 1, 2] + 
               18 I[0, 5, 1, 3] - 6 I[0, 5, 1, 4] - 
               16 I[0, 5, 2, 1] + 31 I[0, 5, 2, 2] - 
               14 I[0, 5, 2, 3] \nonumber\\
               &- I[0, 5, 2, 4] + 14 I[0, 5, 3, 1] - 
               12 I[0, 5, 3, 2] - 2 I[0, 5, 3, 3] - 4 I[0, 5, 4, 1] - 
               I[0, 5, 4, 2] + 5 I[1, 4, 2, 2] \nonumber\\
               &- 10 I[1, 4, 2, 3] + 
               5 I[1, 4, 2, 4] - 10 I[1, 4, 3, 2] + 
               10 I[1, 4, 3, 3] + 5 I[1, 4, 4, 2]\big)\Big)\Bigg)
               \nonumber\\
               &+ 
    e_s \Bigg (-160 m_c^2 P_ 1 \Big (I[0, 3, 1, 0] + I[0, 3, 1, 1] - 
          3 I[0, 3, 1, 2] + I[0, 3, 1, 3] - 2 I[0, 3, 2, 0] + 
          I[0, 3, 3, 0] - I[0, 3, 3, 1] \nonumber\\
          &+ 3 I[1, 2, 1, 1] - 
          3 I[1, 2, 1, 3] - 6 I[1, 2, 2, 1] + 3 I[1, 2, 3, 1]\Big) + 
       5 P_ 1 \Big (25 I[0, 4, 2, 0] - 75 I[0, 4, 2, 1]+ 
          75 I[0, 4, 2, 2] \nonumber\\
       & - 25 I[0, 4, 2, 3] - 71 I[0, 4, 3, 0] + 
          142 I[0, 4, 3, 1] - 71 I[0, 4, 3, 2] + 67 I[0, 4, 4, 0] - 
          67 I[0, 4, 4, 1] - 21 I[0, 4, 5, 0] \nonumber\\
          & + 268 I[1, 3, 2, 1] - 
          284 I[1, 3, 2, 2] + 100 I[1, 3, 2, 3] - 536 I[1, 3, 3, 1] + 
          284 I[1, 3, 3, 2] + 268 I[1, 3, 4, 1]\Big) \nonumber\\
          &+ 
       960 \pi^2 m_c P_ 2  \Big (2 P_ 1 \big (I[0, 2, 1, 0] - 
             2 I[0, 2, 1, 1] + I[0, 2, 1, 2] - 2 I[0, 2, 2, 0] + 
             2 I[0, 2, 2, 1] + I[0, 2, 3, 0] \nonumber\\
             & - 2 I[1, 1, 1, 0] + 
             4 I[1, 1, 1, 1] - 2 I[1, 1, 1, 2] + 4 I[1, 1, 2, 0] - 
             4 I[1, 1, 2, 1] - 2 I[1, 1, 3, 0]\big)- 
          9 \big (I[0, 4, 2, 0] \nonumber\\
             & - 3 I[0, 4, 2, 1] + 3 I[0, 4, 2, 2] - 
              I[0, 4, 2, 3] - 3 I[0, 4, 3, 0] + 6 I[0, 4, 3, 1] - 
              3 I[0, 4, 3, 2] + 3 I[0, 4, 4, 0]  \nonumber\\
              & - 3 I[0, 4, 4, 1] - 
              I[0, 4, 5, 0]+ 12 I[1, 3, 2, 1] - 12 I[1, 3, 2, 2] + 
              4 I[1, 3, 2, 3] - 24 I[1, 3, 3, 1] + 12 I[1, 3, 3, 2] \nonumber\\
              &+ 
              12 I[1, 3, 4, 1]\big)\Big) - 
       324 \Big (I[0, 6, 3, 0] - 4 I[0, 6, 3, 1] + 6 I[0, 6, 3, 2] - 
           4 I[0, 6, 3, 3] + I[0, 6, 3, 4] - 3 I[0, 6, 4, 0] \nonumber\\
           &+ 
           9 I[0, 6, 4, 1] - 9 I[0, 6, 4, 2] + 3 I[0, 6, 4, 3] + 
           3 I[0, 6, 5, 0] - 6 I[0, 6, 5, 1] + 3 I[0, 6, 5, 2] - 
           I[0, 6, 6, 0] + I[0, 6, 6, 1] \nonumber\\
           &+ 6 I[1, 5, 3, 1] - 
           18 I[1, 5, 3, 2] + 18 I[1, 5, 3, 3] - 6 I[1, 5, 3, 4] - 
           18 I[1, 5, 4, 1] + 36 I[1, 5, 4, 2] - 18 I[1, 5, 4, 3]\nonumber\\
           &+ 
           18 I[1, 5, 5, 1] - 18 I[1, 5, 5, 2] - 
           6 I[1, 5, 6, 1]\Big)\Bigg) \nonumber\\
            &+ 
    e_u \Bigg (-160 m_c^2 P_ 1 \Big (16 \pi^2 m_s P_ 3  (I[0, 1, 1, 
               0] - I[0, 1, 1, 1] - I[0, 1, 2, 0]) - I[0, 3, 1, 0] - 
           I[0, 3, 1, 1] + 3 I[0, 3, 1, 2]\nonumber\\
           &- I[0, 3, 1, 3] + 
           2 I[0, 3, 2, 0] - I[0, 3, 3, 0] + I[0, 3, 3, 1] - 
           3 I[1, 2, 1, 1] + 3 I[1, 2, 1, 3] + 6 I[1, 2, 2, 1]- 
           3 I[1, 2, 3, 1]\Big) \nonumber
           \end{align}

\begin{align}
                     & - 
        5 P_ 1 \Big (25 I[0, 4, 2, 0] - 75 I[0, 4, 2, 1] + 
           75 I[0, 4, 2, 2] - 25 I[0, 4, 2, 3] - 71 I[0, 4, 3, 0] + 
           142 I[0, 4, 3, 1] \nonumber\\
           &- 71 I[0, 4, 3, 2] + 67 I[0, 4, 4, 0] - 
           67 I[0, 4, 4, 1] - 21 I[0, 4, 5, 0] + 268 I[1, 3, 2, 1] - 
           284 I[1, 3, 2, 2] + 100 I[1, 3, 2, 3] \nonumber\\
           &- 
           536 I[1, 3, 3, 1] + 284 I[1, 3, 3, 2] + 
           268 I[1, 3, 4, 1]\Big) + 
        48 m_c \bigg (-10 P_ 1 \Big(4 P_ 3 \pi^2 (I[0, 2, 1, 0] - 
                 2 I[0, 2, 1, 1] \nonumber\\
                 &+ I[0, 2, 1, 2] - 2 I[0, 2, 2, 0] + 
                 2 I[0, 2, 2, 1] + I[0, 2, 3, 0] - 2 I[1, 1, 1, 0] + 
                 4 I[1, 1, 1, 1] - 2 I[1, 1, 1, 2] + 
                 4 I[1, 1, 2, 0]\nonumber\\
                 & - 4 I[1, 1, 2, 1] - 
                 2 I[1, 1, 3, 0]) + 
              m_s \big(I[0, 3, 1, 0] - 3 I[0, 3, 1, 1] + 3 I[0, 3, 1, 2] -
                  I[0, 3, 1, 3] - 2 I[0, 3, 2, 0] \nonumber\\
                  &+ 4 I[0, 3, 2, 1] - 
                 2 I[0, 3, 2, 2] + I[0, 3, 3, 0] - I[0, 3, 3, 1] + 
                 3 I[1, 2, 1, 1] - 6 I[1, 2, 1, 2] + 
                 3 I[1, 2, 1, 3] - 6 I[1, 2, 2, 1] \nonumber\\
    %
                   &+ 
                 6 I[1, 2, 2, 2] + 3 I[1, 2, 3, 1]\big)\Big) + 
           9 \Big (20  \pi^2 P_ 3 (I[0, 4, 2, 0] - 3 I[0, 4, 2, 1] + 
                  3 I[0, 4, 2, 2] - I[0, 4, 2, 3] - 3 I[0, 4, 3, 0]\nonumber\\
                   &+ 
                  6 I[0, 4, 3, 1] - 3 I[0, 4, 3, 2] + 
                  3 I[0, 4, 4, 0] - 3 I[0, 4, 4, 1] - I[0, 4, 5, 0] + 
                  12 I[1, 3, 2, 1] - 12 I[1, 3, 2, 2] \nonumber\\
                   &+ 
                  4 I[1, 3, 2, 3] - 24 I[1, 3, 3, 1] + 
                  12 I[1, 3, 3, 2] + 12 I[1, 3, 4, 1]) + 
               3 m_s \big (I[0, 5, 2, 0] - 4 I[0, 5, 2, 1] + 
                   6 I[0, 5, 2, 2] \nonumber\\
                  & - 4 I[0, 5, 2, 3] + 
                   I[0, 5, 2, 4] - 3 I[0, 5, 3, 0] + 
                   9 I[0, 5, 3, 1] - 9 I[0, 5, 3, 2] + 
                   3 I[0, 5, 3, 3] + 3 I[0, 5, 4, 0] \nonumber\\
                   &- 
                   6 I[0, 5, 4, 1] + 3 I[0, 5, 4, 2] - I[0, 5, 5, 0] +
                    I[0, 5, 5, 1] + 5 I[1, 4, 2, 1] - 
                   15 I[1, 4, 2, 2] + 15 I[1, 4, 2, 3]\nonumber\\
                   &- 
                   5 I[1, 4, 2, 4] - 15 I[1, 4, 3, 1] + 
                   30 I[1, 4, 3, 2] - 15 I[1, 4, 3, 3] + 
                   15 I[1, 4, 4, 1] - 15 I[1, 4, 4, 2] - 
                   5 I[1, 4, 5, 1]\big)\Big)\bigg)\nonumber\\
                   &+ 
        324 \Big (I[0, 6, 3, 0] - 4 I[0, 6, 3, 1] + 6 I[0, 6, 3, 2] - 
            4 I[0, 6, 3, 3] + I[0, 6, 3, 4] - 3 I[0, 6, 4, 0] + 
            9 I[0, 6, 4, 1] - 9 I[0, 6, 4, 2]\nonumber\\
            &+ 3 I[0, 6, 4, 3] + 
            3 I[0, 6, 5, 0] - 6 I[0, 6, 5, 1] + 3 I[0, 6, 5, 2] - 
            I[0, 6, 6, 0] + I[0, 6, 6, 1] + 6 I[1, 5, 3, 1]- 
            18 I[1, 5, 3, 2] \nonumber\\
            & + 18 I[1, 5, 3, 3] - 6 I[1, 5, 3, 4] - 
            18 I[1, 5, 4, 1] + 36 I[1, 5, 4, 2] - 18 I[1, 5, 4, 3] + 
            18 I[1, 5, 5, 1] \nonumber\\
            &- 18 I[1, 5, 5, 2] - 
            6 I[1, 5, 6, 1]\Big)\Bigg) 
            +20\,\pi^2\,f_{3\gamma} P_3 m_s 
 \Big(4\,(32e_s+41e_u)\,I[0,3,4,0]+3\,(9e_s+20e_u)\,I[0,3,5,0]\Big)I_2[\mathcal{V}]\nonumber\\
 &+3 m_0^2(e_s+14e_u)\,I_4[\tilde S]\, I[0,5,4,0]\Bigg\},
        \end{align}
  and       
        \begin{align}
         \Delta_4^{QCD}&=\frac {1} {47185920 \pi^5}\bigg\{4  e_c m_c \Bigg (-480 \pi^2 P_ 1 \
(P_ 2 - P_ 3) \Big(I[0, 2, 1, 0] - 2 I[0, 2, 1, 1] + I[0, 2, 1, 2] -           2 I[0, 2, 2, 0]  \nonumber\\
& + 2 I[0, 2, 2, 1]+ I[0, 2, 3, 0] - 
          2 (I[1, 1, 1, 0] - 2 I[1, 1, 1, 1] + I[1, 1, 1, 2] - 
              2 I[1, 1, 2, 0] + 2 I[1, 1, 2, 1] \nonumber\\
              &+ 
              I[1, 1, 3, 0])\Big) + 
       20 m_s P_ 1 \Big(7 I[0, 3, 1, 0] - 20 I[0, 3, 1, 1] + 
          19 I[0, 3, 1, 2] - 6 I[0, 3, 1, 3]- 14 I[0, 3, 2, 0]  \nonumber\\
          &+ 
          26 I[0, 3, 2, 1] - 12 I[0, 3, 2, 2] + 7 I[0, 3, 3, 0] - 
          6 I[0, 3, 3, 1] + 
          18 (I[1, 2, 1, 1] - 2 I[1, 2, 1, 2]+ I[1, 2, 1, 3]\nonumber\\
          &  - 
              2 I[1, 2, 2, 1] + 2 I[1, 2, 2, 2] + 
              I[1, 2, 3, 1])\Big) + 
       360 \pi^2 (P_ 2 - P_ 3) \Big ( -3 I[0, 4, 1, 0]+ 
          9 I[0, 4, 1, 1] - 9 I[0, 4, 1, 2] \nonumber\\
          &+ 3 I[0, 4, 1, 3] + 
          10 I[0, 4, 2, 0]- 21 I[0, 4, 2, 1]+ 12 I[0, 4, 2, 2] - 
          I[0, 4, 2, 3] - 11 I[0, 4, 3, 0] + 13 I[0, 4, 3, 1]\nonumber\\
          &  - 
          2 I[0, 4, 3, 2] + 4 I[0, 4, 4, 0]- I[0, 4, 4, 1] + 
          m_ 0^2 \big(I[0, 3, 2, 0] - 2 I[0, 3, 2, 1] + I[0, 3, 2, 2]- 
             2 I[0, 3, 3, 0] \nonumber\\
          &  + 2 I[0, 3, 3, 1] + I[0, 3, 4, 0] + 
             6 I[1, 2, 2, 1] - 3 I[1, 2, 2, 2]- 6 I[1, 2, 3, 1]\big) + 
          4 (I[1, 3, 2, 1] - 2 I[1, 3, 2, 2] \nonumber\\
             & + I[1, 3, 2, 3] - 
              2 I[1, 3, 3, 1] + 2 I[1, 3, 3, 2] + 
              I[1, 3, 4, 1])\Big) - 
       27 m_s \Big (6 I[0, 5, 1, 1] - 18 I[0, 5, 1, 2] + 
           18 I[0, 5, 1, 3] \nonumber\\
           &- 6 I[0, 5, 1, 4] - 20 I[0, 5, 2, 1] + 
           41 I[0, 5, 2, 2] - 22 I[0, 5, 2, 3] + I[0, 5, 2, 4] + 
           22 I[0, 5, 3, 1] - 24 I[0, 5, 3, 2]\nonumber\\
           &+ 2 I[0, 5, 3, 3] - 
           8 I[0, 5, 4, 1] + I[0, 5, 4, 2] - 
           5 (I[1, 4, 2, 2] - 2 I[1, 4, 2, 3] + I[1, 4, 2, 4] - 
               2 I[1, 4, 3, 2] \nonumber\\
               &+ 2 I[1, 4, 3, 3] + 
               I[1, 4, 4, 2])\Big)\Bigg)\nonumber
                                           \end{align}
               \begin{align}
                &- 
    e_s \Bigg (-15 P_ 1 (I[0, 4, 2, 0] - 3 I[0, 4, 2, 1] + 
          3 I[0, 4, 2, 2] - I[0, 4, 2, 3] - 3 I[0, 4, 3, 0] + 
          6 I[0, 4, 3, 1] - 3 I[0, 4, 3, 2] \nonumber\\
          &+ 3 I[0, 4, 4, 0] - 
          3 I[0, 4, 4, 1] - I[0, 4, 5, 0] + 
          4 (3 I[1, 3, 2, 1] - 3 I[1, 3, 2, 2] + I[1, 3, 2, 3] + 
             3 (-2 I[1, 3, 3, 1] \nonumber\\
             &+ I[1, 3, 3, 2] + I[1, 3, 4, 1]))) + 
       6480 \pi^2 m_c P_ 2  \Big (I[0, 4, 2, 0] - 3 I[0, 4, 2, 1] + 
          3 I[0, 4, 2, 2] - I[0, 4, 2, 3] - 3 I[0, 4, 3, 0] \nonumber\\
          &+ 
          6 I[0, 4, 3, 1] - 3 I[0, 4, 3, 2] + 3 I[0, 4, 4, 0] - 
          3 I[0, 4, 4, 1] - I[0, 4, 5, 0] + 
          4 (3 I[1, 3, 2, 1] - 3 I[1, 3, 2, 2] \nonumber\\
          &+ I[1, 3, 2, 3] + 
              3 (-2 I[1, 3, 3, 1] + I[1, 3, 3, 2] + 
                 I[1, 3, 4, 1]))\Big) + 
       8 m_c^2 \Big (20 P_ 1 (I[0, 3, 1, 0] - 3 I[0, 3, 1, 1]\nonumber\\
                      &+ 
             3 I[0, 3, 1, 2] - I[0, 3, 1, 3] - 2 I[0, 3, 2, 0] + 
             4 I[0, 3, 2, 1] - 2 I[0, 3, 2, 2] + I[0, 3, 3, 0] - 
             I[0, 3, 3, 1] + 
             3 (I[1, 2, 1, 1] \nonumber\\
             &- 2 I[1, 2, 1, 2] + I[1, 2, 1, 3] - 
                2 I[1, 2, 2, 1] + 2 I[1, 2, 2, 2] + I[1, 2, 3, 1])) + 
          27 \Big (2 I[0, 5, 2, 1] - 5 I[0, 5, 2, 2] \nonumber\\
          &+ 
              4 I[0, 5, 2, 3] - I[0, 5, 2, 4] - 4 I[0, 5, 3, 1] + 
              6 I[0, 5, 3, 2] - 2 I[0, 5, 3, 3] + 2 I[0, 5, 4, 1] - 
              I[0, 5, 4, 2] \nonumber\\
              &+ 
              5 (I[1, 4, 2, 2] - 2 I[1, 4, 2, 3] + I[1, 4, 2, 4] - 
                  2 I[1, 4, 3, 2] + 2 I[1, 4, 3, 3] + 
                  I[1, 4, 4, 2])\Big)\Big) + 
       243 \Big (I[0, 6, 3, 0] \nonumber\\
       &- 4 I[0, 6, 3, 1] + 6 I[0, 6, 3, 2] - 
           4 I[0, 6, 3, 3] + I[0, 6, 3, 4] - 3 I[0, 6, 4, 0] + 
           9 I[0, 6, 4, 1] - 9 I[0, 6, 4, 2] \nonumber\\
           &+ 3 I[0, 6, 4, 3] + 
           3 I[0, 6, 5, 0] - 6 I[0, 6, 5, 1] + 3 I[0, 6, 5, 2] - 
           I[0, 6, 6, 0] + I[0, 6, 6, 1] + 
           6 (I[1, 5, 3, 1] \nonumber\\
           &- 3 I[1, 5, 3, 2] + 3 I[1, 5, 3, 3] - 
               I[1, 5, 3, 4] - 
               3 (I[1, 5, 4, 1] - 2 I[1, 5, 4, 2] + I[1, 5, 4, 3] - 
                  I[1, 5, 5, 1] + I[1, 5, 5, 2])\nonumber\\
                  &- 
               I[1, 5, 6, 1])\Big)\Bigg)\nonumber\\
                &+ 
    e_u \Bigg (-15 P_ 1 \Big (I[0, 4, 2, 0] - 3 I[0, 4, 2, 1] + 
           3 I[0, 4, 2, 2] - I[0, 4, 2, 3] - 3 I[0, 4, 3, 0] + 
           6 I[0, 4, 3, 1] - 3 I[0, 4, 3, 2]\nonumber\\
           &+ 3 I[0, 4, 4, 0] - 
           3 I[0, 4, 4, 1] - I[0, 4, 5, 0] + 
           4 (3 I[1, 3, 2, 1] - 3 I[1, 3, 2, 2] + I[1, 3, 2, 3] + 
               3 (-2 I[1, 3, 3, 1] \nonumber\\
               &+ I[1, 3, 3, 2] + 
                  I[1, 3, 4, 1]))\Big) + 
        8 m_c^2 \Big (-720 \pi^2 m_s P_ 3  \Big (I[0, 3, 2, 0] - 
              2 I[0, 3, 2, 1] + I[0, 3, 2, 2] - 2 I[0, 3, 3, 0]\nonumber\\
              &+ 
              2 I[0, 3, 3, 1] + I[0, 3, 4, 0] + 6 I[1, 2, 2, 1] - 
              3 I[1, 2, 2, 2] - 6 I[1, 2, 3, 1]\Big) + 
           20 P_ 1 \Big (I[0, 3, 1, 0] - 3 I[0, 3, 1, 1] \nonumber\\
           &+ 
              3 I[0, 3, 1, 2] - I[0, 3, 1, 3] - 2 I[0, 3, 2, 0] + 
              4 I[0, 3, 2, 1] - 2 I[0, 3, 2, 2] + I[0, 3, 3, 0] - 
              I[0, 3, 3, 1] + 
              3 (I[1, 2, 1, 1] \nonumber\\
              &- 2 I[1, 2, 1, 2] + I[1, 2, 1, 3] - 
                  2 I[1, 2, 2, 1] + 2 I[1, 2, 2, 2] + 
                  I[1, 2, 3, 1])\Big) + 
           27 \Big (2 I[0, 5, 2, 1] - 5 I[0, 5, 2, 2] \nonumber\\
           &+ 
               4 I[0, 5, 2, 3] - I[0, 5, 2, 4] - 4 I[0, 5, 3, 1] + 
               6 I[0, 5, 3, 2] - 2 I[0, 5, 3, 3] + 2 I[0, 5, 4, 1] - 
               I[0, 5, 4, 2] \nonumber\\
               &+ 
               5 (I[1, 4, 2, 2] - 2 I[1, 4, 2, 3] + I[1, 4, 2, 4] - 
                   2 I[1, 4, 3, 2] + 2 I[1, 4, 3, 3] + 
                   I[1, 4, 4, 2])\Big)\Big) + 
        324 m_c \Big (20  \pi^2 P_ 3  (I[0, 4, 2, 0] \nonumber\\
        &- 
              3 I[0, 4, 2, 1] + 3 I[0, 4, 2, 2] - I[0, 4, 2, 3] - 
              3 I[0, 4, 3, 0] + 6 I[0, 4, 3, 1] - 3 I[0, 4, 3, 2] + 
              3 I[0, 4, 4, 0] - 3 I[0, 4, 4, 1]\nonumber\\
              &- I[0, 4, 5, 0] + 
              4 (3 I[1, 3, 2, 1] - 3 I[1, 3, 2, 2] + I[1, 3, 2, 3] + 
                 3 (-2 I[1, 3, 3, 1] + I[1, 3, 3, 2] + 
                    I[1, 3, 4, 1]))) \nonumber\\
                    &+ 
           3 m_s \Big (I[0, 5, 2, 0] - 4 I[0, 5, 2, 1] + 
               6 I[0, 5, 2, 2] - 4 I[0, 5, 2, 3] + I[0, 5, 2, 4] - 
               3 I[0, 5, 3, 0] + 9 I[0, 5, 3, 1] - 9 I[0, 5, 3, 2]\nonumber\\
               &+ 
               3 I[0, 5, 3, 3] + 3 I[0, 5, 4, 0] - 6 I[0, 5, 4, 1] + 
               3 I[0, 5, 4, 2] - I[0, 5, 5, 0] + I[0, 5, 5, 1] + 
               5 (I[1, 4, 2, 1] - 3 I[1, 4, 2, 2]\nonumber\\
               &+ 3 I[1, 4, 2, 3] - 
                   I[1, 4, 2, 4] - 
                   3 (I[1, 4, 3, 1] - 2 I[1, 4, 3, 2] + 
                    I[1, 4, 3, 3] - I[1, 4, 4, 1] + I[1, 4, 4, 2]) - 
                   I[1, 4, 5, 1])\Big)\Big) \nonumber\\
                   &+ 
        243 \Big (I[0, 6, 3, 0] - 4 I[0, 6, 3, 1] + 6 I[0, 6, 3, 2] - 
            4 I[0, 6, 3, 3] + I[0, 6, 3, 4] - 3 I[0, 6, 4, 0] + 
            9 I[0, 6, 4, 1] - 9 I[0, 6, 4, 2] \nonumber\\
            &+ 3 I[0, 6, 4, 3] + 
            3 I[0, 6, 5, 0] - 6 I[0, 6, 5, 1] + 3 I[0, 6, 5, 2] - 
            I[0, 6, 6, 0] + I[0, 6, 6, 1] + 
            6 (I[1, 5, 3, 1] - 3 I[1, 5, 3, 2] \nonumber\\
            &+ 3 I[1, 5, 3, 3] - 
                I[1, 5, 3, 4] - 
                3 (I[1, 5, 4, 1] - 2 I[1, 5, 4, 2] + I[1, 5, 4, 3] - 
                   I[1, 5, 5, 1] + I[1, 5, 5, 2]) - 
                I[1, 5, 6, 1])\Big)\Bigg)\nonumber\\
                &+m_0^2\Big(-120\, (e_s + 10 e_u)\, f_{3\gamma}\, \pi^2\, I_2[\mathcal V]\, I[0, 4, 4, 0]
+ (3 e_s + 14 e_u) I_4[\mathcal S]\, I[0, 5, 4, 0]+ 2 e_s\, I_4[\mathcal T_1] I[0, 5, 4, 0]\Big)\Bigg\},
        \end{align}
where $P_1 =\langle g_s^2 G^2\rangle$, $P_2 =\langle \bar q q \rangle$ and $P_3 = \langle \bar s s \rangle$ are gluon, u/d and s-quark condensates, respectively. It should be noted that in the above terms, for the sake of simplicity, we only present expressions that make significant contributions to the numerical values of the magnetic moments and do not give much higher dimensional contributions, although they are taken into account in numerical computations.

The functions~$I[n,m,l,k]$, $I_1[\mathcal{A}]$,~$I_2[\mathcal{A}]$,~$I_3[\mathcal{A}]$,~$I_4[\mathcal{A}]$,
~$I_5[\mathcal{A}]$, and ~$I_6[\mathcal{A}]$ are
defined as:
\begin{align}
 I[n,m,l,k]&= \int_{4 m_c^2}^{s_0} ds \int_{0}^1 dt \int_{0}^1 dw~ e^{-s/M^2}~
 s^n\,(s-4\,m_c^2)^m\,t^l\,w^k,\nonumber\\
 I_1[\mathcal{A}]&=\int D_{\alpha_i} \int_0^1 dv~ \mathcal{A}(\alpha_{\bar q},\alpha_q,\alpha_g)
 \delta'(\alpha_ q +\bar v \alpha_g-u_0),\nonumber\\
  I_2[\mathcal{A}]&=\int D_{\alpha_i} \int_0^1 dv~ \mathcal{A}(\alpha_{\bar q},\alpha_q,\alpha_g)
 \delta'(\alpha_{\bar q}+ v \alpha_g-u_0),\nonumber\\
   I_3[\mathcal{A}]&=\int D_{\alpha_i} \int_0^1 dv~ \mathcal{A}(\alpha_{\bar q},\alpha_q,\alpha_g)
 \delta(\alpha_ q +\bar v \alpha_g-u_0),\nonumber\\
   I_4[\mathcal{A}]&=\int D_{\alpha_i} \int_0^1 dv~ \mathcal{A}(\alpha_{\bar q},\alpha_q,\alpha_g)
 \delta(\alpha_{\bar q}+ v \alpha_g-u_0),\nonumber\\
   I_5[\mathcal{A}]&=\int_0^1 du~ A(u)\delta'(u-u_0),\nonumber\\
 I_6[\mathcal{A}]&=\int_0^1 du~ A(u),\nonumber
 \end{align}
 where $\mathcal{A}$ represents the corresponding photon DAs.

\end{widetext}

\bibliography{ZcstetraquarksN}

\end{document}